\documentclass[traditabstract]{aa}
 \pdfoutput=1
\newcommand{\lsim}{\lower 2pt \hbox{$\, \buildrel {\scriptstyle
<}\over {\scriptstyle \sim}\,$}}  \newcommand{\gsim}{\lower 2pt
\hbox{$\, \buildrel {\scriptstyle >}\over {\scriptstyle \sim}\,$}}
\usepackage{appendix}
\usepackage{amsmath}
\usepackage{graphicx}
\usepackage{txfonts}
\usepackage{natbib}
\usepackage{longtable}
\usepackage{rotating}
\usepackage{lscape}
\usepackage{float}
\usepackage{subfig}
\usepackage{caption}
\usepackage{multirow}
\usepackage{array}
\usepackage{fixltx2e}
\usepackage{hyperref}
\usepackage{graphicx}
\usepackage{txfonts}

\bibpunct{(}{)}{;}{a}{}{,}

\newcommand{\Pa}{Pa$\alpha$}
\newcommand{\Brd}{Br$\delta$}
\newcommand{\Brg}{Br$\gamma$}

\newcommand{\LTIR}{L$_{\rm IR}$}
\newcommand{\Lsolar}{L$_{\odot}$}

\newcommand{\Av}{A$_{\rm V}$}

\newcommand\nodata{ ~$\cdots$~ }

\newcolumntype{x}[1]{%
>{\raggedleft\hspace{0pt}}p{#1}}%

\newcolumntype{z}[1]{%
>{\raggedright\hspace{0pt}}p{#1}}%

\begin{document}

   \title{Understanding the two-dimensional ionization structure in luminous infrared galaxies.}

   \subtitle{A near-IR integral field spectroscopy perspective }

   \author{L. Colina\inst{1} \and J. Piqueras L\'opez\inst{1} \and S. Arribas\inst{1} \and Rogerio Riffel\inst{2}
    \and Rogemar A. Riffel\inst{3} \and A. Rodriguez-Ardila\inst{4} \and M. Pastoriza\inst{2} \and 
    T. Storchi-Bergmann\inst{2} \and A. Alonso-Herrero\inst{5} \and D. Sales\inst{2}
          }
   \institute{Centro de Astrobiolog\'ia (CAB, CSIC-INTA), Carretera de Ajalvir,
              Torrej\'on de Ardoz, Madrid\\
              \email{colina@cab.inta-csic.es}
         \and
             Universidade Federal Rio Grande do Sul (UFRGS), RS, Brazil
\and
             Departamento de F\'isica, Centro de Ciencias Naturais e Exatas, 
             Universidade Federal de Santa Maria (UFSM), 97105-900, Santa Mar\'ia, RS, Brazil 
\and
            Laboratorio Nacional de Astrofisica, Itajub\'a, MG, Brazil
\and
            Instituto de F\'isica de Cantabria, Santander, Spain
             }

\date{}

\abstract{We investigate the two-dimensional excitation structure of the interstellar medium in a sample of luminous infrared galaxies (LIRGs) and Seyferts using near-IR integral field spectroscopy. This study extends to the near infrared the well-known optical and mid-IR emission line diagnostics used to classify activity in galaxies. Based on the spatially resolved spectroscopy of prototypes, we identify in the [FeII]1.64$\mu$m/Br$\gamma$ $-$ H$_2$2.12$\mu$m/Br$\gamma$ plane  regions dominated by the different heating sources, i.e. AGNs, young main-sequence massive stars, and evolved stars i.e. supernovae.  The ISM in LIRGs occupy a wide region in the near-IR diagnostic plane from $-$0.6 to $+$1.5 and from $-$1.2 to $+$0.8 (in log units) for the [FeII]/Br$\gamma$ and H$_2$/Br$\gamma$ line ratios, respectively. The corresponding median(mode) ratios are +0.18(0.16) and +0.02($-$0.04). Seyferts show on average larger values by factors $\sim$ 2.5 and $\sim$1.4 for the [FeII]/Br$\gamma$ and H$_2$/Br$\gamma$ ratios, respectively.

New areas and relations in the near-IR diagnostic plane are defined for the compact, high surface brightness regions dominated by AGN, young ionizing stars, and supernovae explosions, respectively.
In addition to these high surface brightness regions, the diffuse regions affected by the AGN radiation field cover an area similar to that of Seyferts, but with high values in [FeII]/Br$\gamma$ that are not as extreme. The extended, non-AGN diffuse regions cover a wide area in the near-IR diagnostic diagram that overlaps  that of  individual excitation mechanisms (i.e. AGN, young stars, and supernovae), but with its mode value to that of the young star-forming clumps. This indicates that the excitation conditions of the extended, diffuse ISM are likely due to a mixture of the different ionization sources, weighted by their spatial distribution and relative flux contribution.  The integrated line ratios in LIRGs show higher excitation conditions i.e. towards AGNs, than those measured by the spatially resolved spectroscopy. If this behaviour is representative, it would have clear consequences when classifying high-z, star-forming galaxies based on their near-infrared integrated spectra.}


\maketitle

\section{Introduction}

The characterization of the excitation and ionization conditions of the interstellar medium plays a central role in our understanding of galaxy formation and evolution across the history of the Universe. Galaxies are known to have different excitation/heating sources ranging from luminous accretion disks around nuclear massive black holes (AGNs), to powerful, spatially distributed starbursts. In general, galaxies show a complex excitation/ionization, two-dimensional structure due to their different radiation field (stars versus AGNs), the presence of local (radio jets, stellar and supernovae winds) or galaxy-wide (tidally-induced) shocks, internal dust clumpiness, and metallicity effects (radial gradients, overall abundances). Thus, the use of strong emission lines for the detection and quantification of the ionization sources in galaxies was identified early on as an important tool. The pioneering methodology of Baldwin, Phillips, and Terlevich (BPT; \citealt{Baldwin:1981PASP93}), based on the use of the strongest optical emission lines, has been commonly used ever since to identify and separate AGN from star formation (\citealt{Veilleux1987ApJS63}, \citealt{Kewley:2001p8174}, \citealt{Yuan:2010eo}) to define empirical limits based on large samples of galaxies \citep{Kauffmann:2003MNRAS346} or theoretical limits based on detailed models (\citealt{Kewley:2006p5048}, \citeyear{Kewley:2013cs}). The advent of infrared satellites like ISO and Spitzer has also made it possible to define BPT-like diagnostics (\citealt{Genzel:1998ApJ498}, \citealt{Spoon:2007ApJ654L}, \citealt{Groves:2008ij}, \citealt{BernardSalas:2009ApJS184}, \citealt{Dale:2009p6943}, \citealt{PereiraSantaella:2011fw}, \citealt{Petric:2011ApJ730}, \citealt{AlonsoHerrero:2012p744}) based on strong mid-IR lines that are tracers of AGNs (i.e. high excitation lines like [NeV]14.32$\mu$m or [OIV]25.89$\mu$m) and star formation (PAHs at 6.2, 7.7 and 11.3 $\mu$m). 

The study of the excitation and ionization of ISM in active galaxies using near-IR emission lines has been more limited. A BPT-like near-IR diagram involving the [FeII]1.26$\mu$m/Pa$\beta$ and the H$_2$2.12$\mu$m/Br$\gamma$ line ratios was originally proposed by \cite{Larkin:1998p5053} in a study of LINERs based on low spectral resolution (R $\sim$ 1000) near-IR long-slit spectroscopy. Larkin and coworkers reported a strong linear relation in the (log$-$log) [FeII]1.26$\mu$m/Pa$\beta$ $-$ H$_2$2.12$\mu$m/Br$\gamma$ plane, with star-forming galaxies showing the lowest ratios, Seyferts with intermediate ratios, and LINERs with the highest ratios. Later studies of Seyfert 1 and Seyfert 2 galaxies (\citealt{Rodriguez-Ardila:2004p425}, \citeyear{Rodriguez-Ardila:2005p364}, \citealt{Riffel:2006A&A457}) confirmed the trend first reported by Larkin and coworkers. A more recent analysis on a larger compilation of Seyfert 1 and 2 galaxies, including star-forming galaxies and a few LINERs \citep{Riffel:2013MNRAS429}, have shown that activity in galaxies seems to follow a unique linear relation log([FeII]1.26$\mu$m/Pa$\beta$)= 0.749$\times$log(H$_2$2.12$\mu$m/Br$\gamma$) $-$ 0.207, with star-forming galaxies showing the lowest ratios and LINERs apparently the largest.       

The ionization of the ISM in luminous and ultraluminous infrared galaxies (LIRGs and ULIRGs, respectively) shows a complex behaviour in the optical with the largest fraction of galaxies classified as star forming, a large fraction (30\% to 40\%) as LINERs, and a few as AGNs (\citealt{Veilleux:1995p98}, \citealt{Veilleux:1999p7887}). The nature of LINERs in U/LIRGs has been explained as the consequence of galaxy-wide merger-driven shocks (\citealt{MonrealIbero:2006p2309}, \citeyear{Monreal-Ibero:2010p517}, \citealt{Rich:2011ApJ734}, \citeyear{Rich:2013ASPC477}), but other alternatives like a mixture of starbursts and AGN have also been invoked (\citealt{Kewley:2001p8174}, \citealt{Yuan:2010eo}, \citealt{Kewley:2013cs}). The study of U/LIRGs in the optical includes additional complications derived by the amount and clumpy nature of their dust distribution, and by the presence of additional heating mechanisms due to the tidal forces associated with the interactions and mergers involved mostly in ULIRGs. 

A more appropriate methodology for the study of the ionization/excitation conditions of the ISM in U/LIRGs could be the use of near-infrared spatially resolved spectroscopy, i.e near-IR integral field spectroscopy (IFS). This allows the investigation of  the low-A$_V$, transparent ionized regions (as in the optical) and the more dust-enshrouded star-forming regions and/or obscured AGNs. These type of studies will set the ground for future studies with the near-IR (NIRSpec) and mid-IR (MIRI) integral field spectrographs onboard the James Webb Space Telescope (JWST). Future observatories such as JWST will allow us  to spatially resolve the high-redshift, infrared-luminous, star-forming galaxies (main sequence and above), establishing and quantifying the nature of their heating/ionizing sources.

This paper presents the first spatially resolved near-infrared spectroscopic study of the excitation conditions of the interstellar medium in a sample of low-redshift LIRGs complemented with similar data from a sample of nearby Seyfert galaxies. Section $\S$2 introduces the samples and near-IR integral field spectroscopic data. The two-dimensional distribution of the ISM in LIRGs according to their [FeII]/Br$\gamma$ and H$_2$/Br$\gamma$ ratios is given in sections $\S$3.1, 3.2 and $\S$3.3.1. The location of compact young star-forming regions, SNe- and AGN-dominated, and composite starbursts+AGN in the [FeII]/Br$\gamma$ $-$ H$_2$/Br$\gamma$ plane are based on galaxies identified as prototypes, and is presented in sections $\S$3.3.2 to $\S$3.3.5. A discussion about possible aperture effects in the classification of galaxies when comparing flux-weighted one-dimensional  with two-dimensional spatially distributed classifications is given ($\S$3.3.6). Finally, new limits and relations in the (log$-$log) [FeII]/Br$\gamma$ versus H$_2$/Br$\gamma$ plane are proposed to discriminate the different excitation associated with young and aged stars, AGNs, and composite objects ($\S$3.3.7).

\section{Samples of galaxies and observations}

\subsection{Luminous infrared galaxies}

\begin{table}[t]
\caption{LIRG sample}
\tiny
\centering
{\setlength{\tabcolsep}{1.8pt}
\begin{tabular}{cccccc}
\hline
\hline
     ID1      &   ID2   & z &   D$_{\rm L}$       &     Scale    & log (\LTIR/\Lsolar) \\
Common  &   IRAS  &    &(Mpc)    &(pc/") &  \\
     (1)       &   (2)     &    (3)   &    (4)      &      (5)     &        (6)      \\
\hline
\object{NGC 2369} & \object{IRAS 07160-6215} & 0.010807& 48.6 & 230 & 11.17 \\
\object{NGC 3110} & \object{IRAS 10015-0614} & 0.016858& 78.4 & 367 & 11.34 \\
\object{NGC 3256} & \object{IRAS 10257-4338} & 0.009354 & 44.6 & 212 & 11.74 \\
\object{ESO 320-G030} & \object{IRAS 11506-3851} & 0.010781& 51.1 & 242 & 11.35\\
\object{IRASF 12115-4656} & \object{IRAS 12115-4657} & 0.018489 & 84.4 & 394 & 11.10 \\
\object{NGC 5135} & \object{IRAS 13229-2934} & 0.013693& 63.5 & 299 & 11.33 \\
\object{IRASF 17138-1017} & \object{IRAS 17138-1017} & 0.017335& 75.3 & 353 & 11.42 \\
\object{IC 4687} & \object{IRAS 18093-5744} & 0.017345 & 75.1 & 352 & 11.44 \\
\object{NGC 7130} & \object{IRAS 21453-3511} & 0.016151 & 66.3 & 312 & 11.34 \\
\object{IC 5179} & \object{IRAS 22132-3705} & 0.011415& 45.6 & 216 & 11.12 \\
\hline
\hline
\end{tabular}}
\tablefoot{Col. (3): redshift from the NASA Extragalactic Database (NED). Cols. (4) and (5): luminosity distance and scale derived using Ned Wright's Cosmology Calculator \citep{Wright:2006p4236} given h$_{0}$ = 0.70, $\Omega_{\rm M}$ = 0.7, $\Omega_{\rm M}$ = 0.3. Col. (6): \LTIR (8--1000$\mu$m) calculated from the IRAS flux densities $f_{12}$, $f_{25}$, $f_{60}$ and $f_{100}$ \citep{Sanders:2003p1433}, using the expression given in \cite{Sanders:1996p845}. A complete version of this table can be found in \cite{Piqueras2012A&A546A}}
\label{table:sample}
\end{table}

The sample of galaxies (see Table~\ref{table:sample}) consists of ten low-z LIRGs located at distances of $\sim$ 45 to 85  Mpc, and covering the IR-luminosity range of 11.1 to 11.7 L$_{\odot}$ (logL$_{IR}$). Most galaxies are classified as pure star forming, while only two (NGC 5135 \& NGC 7130) are identified as hosting  an AGN in their nucleus in addition. We obtained spectra for all galaxies  with the near-infrared integral field spectrograph SINFONI on the VLT, using the seeing-limited mode, and a resolving power of about 3000 and 4000 for the H and K bands, respectively. For each galaxy, we generated absolute calibrated maps  for the H$_2$ 2.12$\mu$m, Br$\gamma$, and [FeII] 1.64$\mu$m emission lines, tracing the distribution of the hot molecular, warm ionized, and partially ionized gas, respectively. The overall field-of-view ($\sim$ 8 $\times$ 8 arcsec$^2$) covers distances of 1 to 1.5 kpc radius around the nucleus of the galaxy with a sampling of about 30 to 50 parsec (0.125 arcsec/spaxel), and a physical resolution of about 200 parsec (FWHM $\sim$ 0.6 arcsec). Full details of the sample observations,  basic reduction, and calibration can be found in \cite{Piqueras2012A&A546A}. 

We included some improvements in the reduction and calibration process of our SINFONI data, as explained in detail elsewhere (Piqueras L\'opez et al. 2015, in prep.). We implemented different techniques to reduce the spectral and spatial noise of the individual observations, and to match the background of the individual cubes before coadding the final data cube. After the new calibration of the data set, we obtained new emission line maps, following the same procedure as described in \cite{Piqueras2012A&A546A}. As a consequence of the new reduction process, the relative calibration between H- and K-band data cubes might be slightly different when compared with previous published measurements. In the worst case scenario, the differences between the 2012 absolute calibration and the calibration that we present  could be as high as 20\% on each band. However, on average, the uncertainty of the absolute flux calibration lies below $\leq15$\% on both bands. Therefore, we assumed a conservative 15\% systematic error for all flux measurements, added in quadrature to the statistical error.

To obtain extinction-corrected flux maps, we applied the spatially resolved internal \Av\ measurements from \cite{PiquerasLopez:2013hx}. These measurements are derived from the \Brg/\Brd\ line ratio and the extinction law from \cite{Calzetti:2000p2349}, and are available on a spaxel-by-spaxel basis for those spaxels where the \Brd\ line is detected beyond a S/N threshold of 4 (see \citealt{PiquerasLopez:2013hx} for further details). In those spaxels where the correction is not available, the median value of the \Av\ distribution is used for each individual galaxy. Strictly speaking, the use of Calzetti's law has to be applied by deriving  the A$_V$ for the stellar continuum first  and then the extinction towards the ionized gas using the empirically (UV-optical) derived ratio \Av(continuum)/\Av(ionized gas)= 0.44. However, while  extinction laws differ in the ultraviolet and optical, there are similar extinction laws in the near infrared. For example, the differences (in magnitudes) between \cite{Cardelli:1989ApJ345} and Calzetti's laws for the wavelengths of the Pa$\alpha$, Br$\delta,$ and Br$\gamma$ emission lines correspond to 0.00193, 0.00658, and 0.02037$\times$\Av\, respectively. 

We obtained integrated measurements of the line ratios on different regions of interest, i.e. the peaks of the emission lines, [FeII]1.64$\mu$m, H$_2$ 1-0S(1) and \Brg, the main nucleus of the objects, and integrated measurements over the field of view (FoV). The peak of each line is defined as the brightest spaxel on the corresponding emission line map. For consistency with our previous work, we consider the nucleus of the galaxy as the brightest spaxel of the K-band continuum map. Centred on the peaks, we defined a circular aperture of 200\,pc radius to obtain an integrated spectrum of each region.  The spaxel-by-spaxel spectra within the aperture are \textup{\emph{\textup{de-rotated}}} i.e. shifted to the same reference frame in the spectral axis before staking (Piqueras L\'opez et al. 2015, in prep.). This procedure increases the S/N of the lines by de-correlating the noise in the spectral dimension, since possible residuals from the sky line subtraction are no longer aligned. It also corrects from the artificial line broadening due to the intrinsic velocity field that arises from staking over wide apertures. After the spectra are extracted, we  individually fitted the different emission lines (i.e. [FeII], H$_2$, \Brg\ and \Brd) using a Gaussian profile on a spaxel-by-spaxel basis (see \citealt{Piqueras2012A&A546A} for details), and obtained the corresponding line fluxes and \Av\ measurements using the \Brg/\Brd\ line ratio. Examples of spectra can be found in \cite{Piqueras2012A&A546A}, The [FeII]1.644$\mu$m emission line flux could  potentially be affected by the Br12 1.641$\mu$m line in absorption. However, these two lines are well resolved in our R $\sim$ 4000 spectra and therefore no major impact on the strong [FeII] emission is expected (see \citealt{Bedregal:2009p2426} for examples of the H-band spectra for NGC 5135).

The final uncertainties on the line fluxes and extinction measurements are calculated using a \emph{\textup{bootstrap}} method of typically $\sim500$ simulations of each spectrum. This method consists of constructing a set of spectra that are statistically equivalent in terms of noise to the observed spectrum. We measured the noise of each observed spectrum as the rms of the residuals after subtraction of the Gaussian profile at each line position. Taking this estimation of the noise into account, we constructed a total of $\sim500$ simulated spectra where the lines are  fitted again. We then obtained distributions of each parameter of the fitting (e.g. flux, line position/velocity, and $\sigma$), and calculated their corresponding uncertainties as the standard deviation of each distribution. The main advantage of this technique is that it not only takes the photon noise into account, but also uncertainties due to an inaccurate continuum estimation or poor line fitting in low S/N spectra.

\subsection{Nearby Seyfert galaxies}

\begin{table}[t]
\caption{Seyfert and star-forming galaxy samples}
\tiny
\centering
{\setlength{\tabcolsep}{1.8pt}
\begin{tabular}{ccccccc}
\hline
\hline
     ID1      &   ID2   & z &   D$_{\rm L}$       &     Scale   \\
Common  &   IRAS   &    &(Mpc)    &(pc/") &  \\
     (1)       &   (2)     &    (3)   &    (4)      &      (5)     \\
\hline
\object{Mrk 1157} & \object{IRAS 01306+3524}  & 0.015167& 61.8 & 291  \\
\object{Mrk 1066} & \object{IRAS 02568+3637}  & 0.012025& 49.2 & 233  \\
\object{ESO 428-G014} & \object{IRAS 07145-2914}  & 0.005664& 27.1 & 130  \\
\object{Mrk 79} & \object{IRAS 07388+4955}  & 0.022189& 98.3 & 456  \\
\object{NGC 4151} & \object{IRAS Z12080+3940}  & 0.003319& 17.8 & 86  \\
\hline
\object{NGC 693} & \object{IRAS 01479+0553}  & 0.005227& 18.3 & 88  \\
\object{UGC 2855} & \object{IRAS 03431+6958}  & 0.004003& 15.9 & 77  \\
\object{NGC 1569W} & \object{IRAS 04260+6444}  & -0.000347 & 1.45 & 7  \\
\object{M82} & \object{IRAS 09517+6954}  & 0.000677& 4.03 & 19  \\
\object{NGC 4303} & \object{IRAS 12194+0444}  & 0.005224& 27.5 & 131  \\
\object{Mrk 59} & \object{IRAS 12566+3507}  & 0.002616& 14.7 & 71  \\
\object{NGC 7771} & \object{IRAS 23488+1949}  & 0.014267& 56.6 & 267  \\
\hline
\hline
\end{tabular}}
\tablefoot{Col. (3): redshift from the NASA Extragalactic Database (NED). Cols. (4) and (5): Luminosity distance and scale derived using Ned Wright's Cosmology Calculator \citep{Wright:2006p4236} given h$_{0}$ = 0.70, $\Omega_{\rm M}$ = 0.7, $\Omega_{\rm M}$ = 0.3.}
\label{table:seyfert}
\end{table}

The LIRG sample has been complemented with a small sample of nearby Seyfert galaxies (see Table~\ref{table:seyfert}) for which high spectral resolution (R $\sim$ 5000) near-IR IFS data obtained with NIFS/Gemini are already available. The sample consists of three Seyfert 2 galaxies (Mrk 1066 \& Mrk 1157; \citealt{Riffel:2011ej};\citealt{Riffel:2011MNRAS417}; ESO428-G014, \citealt{Riffel:2006A&A457}), one Seyfert 1.5 galaxy (NGC 4151, \citealt{StorchiBergmann:2009if}), and one Seyfert 1 galaxy (Mrk 79, \citealt{Riffel:2013ft}). These data cover regions of 300 to 700 parsec (radius) around the AGN with an angular resolution from about 10 parsec (ESO428$-$G014 and NGC 4151) up to about 100 parsec (Mrk 79). One of the Seyferts (Mrk 1066), while dominated by the AGN in the central regions, also has circumnuclear star-forming knots within the central 400 pc around the nucleus (\citealt{Riffel:2011ej}; \citealt{RamosAlmeida:2014gg}).

The [FeII]1.64$\mu$m/\Brg\ line ratios for the Seyfert sample are derived from the original [FeII]$\lambda$1.26\,$\mu$m and Pa$\beta$ line emission maps. To calculate the [FeII]$\lambda$1.64\,$\mu$m/\Brg\ line ratio from the [FeII]$\lambda$1.26\,$\mu$m/\Pa$\beta$, we assumed the theoretical ratio [FeII]$\lambda$1.64\,$\mu$m/[FeII]$\lambda$1.26\,$\mu$m$=0.7646$ (see e.g. \citealt{Colina:1993p6004}), and the case B recombination factors at T=10,000~K and n$_{\rm e} = 10^{4}$\,cm$^{-3}$  \citep{Osterbrock:2006AGN2}. Therefore, [FeII]1.64$\mu$m/Br$\gamma$ = 4.4974 $\times$ [FeII]1.26$\mu$m/Pa$\beta$. Since the emission lines used in the original line ratios lie very closed in wavelength, no correction for internal extinction effects has been applied to the Seyferts.

\subsection{Nearby star-forming galaxies}

There are a few nearby star-forming galaxies with published near-IR IFS. These are included in the analysis for a direct comparison of the excitation properties of the brightest compact emitting regions in LIRGs with the nucleus and star-forming clumps in pure star-forming galaxies. The sample (see Table~\ref{table:seyfert}) includes the nucleus, two off-nucleus regions (35 $\times$ 35 parsec, each), and the central (260 $\times$ 160 parsec) region in M82 \citep{ForsterSchreiber:2001ApJ552}, as well as one low-metallicity blue compact dwarf galaxy (Mrk 59, \citealt{Izotov:2009hi}) and a small sample of nearby star-forming galaxies (NGC 1569W, NGC 693, UGC2855, and NGC 7771; \citealt{Dale:2004ApJ601}), covering several morphologies and metallicities. For all these galaxies, we obtained the [FeII]/\Brg\ and H$_2$/\Brg\ line ratios  from the published line fluxes for the different regions corrected by the specific internal extinction given in these works.

In addition to these galaxies from the literature, the line ratios for the nucleus and a circumnuclear UV-luminous star cluster (cluster G in \citealt{Colina:2002ApJ579}) in the galaxy NGC 4303, located in the Virgo cluster, are also included. The nucleus of this galaxy, optically classified as LINER - low-luminosity Seyfert 2 \citep{Colina:1999ApJ514}, contains a massive (10$^5$ M$\sun$), UV-luminous, 4 Myr old star cluster \citep{Colina:2002ApJ579}, and a hard X-ray emitting low-luminosity AGN \citep{JimenezBailon:2003ApJ593}.

\section{Results}

We generated maps for the observed Br$\gamma$, H$_2$2.12$\mu$m, and [FeII]1.64$\mu$m emission  for each individual LIRG.  Spatial information is typically available for a few to several thousands of spaxels per galaxy with a total number of over 28000 spaxels for all LIRGs in the sample. Since internal extinction is very clumpy in LIRGs and shows a wide range of values \citep{PiquerasLopez:2013hx}, we performed subsequent analysis of the line ratios  after applying an internal extinction correction on a per-spaxel basis  for a large fraction of spaxels (64\%) with a good determination of the A$_V$ based on the Br$\delta$/Br$\gamma$ ratio. For the rest of spaxels in each galaxy, the median value of the corresponding distribution is applied. The results for each LIRG as well as for the entire sample are given in Tables~\ref{table:distrib_fe_LIRGs} \& \ref{table:distrib_h2_LIRGs} for the [FeII]1.64$\mu$m/Br$\gamma$ and H$_2$2.12$\mu$m/Br$\gamma$ ratios, respectively. 

In addition to the maps, we also calculated the [FeII]1.64$\mu$m/Br$\gamma$ and H$_2$2.12$\mu$m/Br$\gamma$ line ratios  in each galaxy for a few high surface brightness regions, mainly the nucleus of the galaxy identified as the photometric peak in the SINFONI K-band line-free continuum, and the emission peaks of the ionized, partially-ionized, and hot molecular gas components as traced by the Br$\gamma$, the [FeII]1.64$\mu$m, and the H$_2$ 2.12$\mu$m lines respectively (see Tables~\ref{table:regions_fe_LIRGs} \& \ref{table:regions_h2_LIRGs}). Finally, the flux-weighted line ratios over the entire FOV are also given for a direct comparison with one-dimensional ratios for large sample of galaxies.

We obtained the corresponding emission line ratios for the sample of Seyfert galaxies  from the original NIFS emission line maps (see $\S$2.2), following the [FeII]1.26$\mu$m/Pa$\beta$ to [FeII]1.64$\mu$m/Br$\gamma$ as explained in $\S$2.2 (see Tables~\ref{table:distrib_fe_Sy} \& \ref{table:distrib_h2_Sy} for results).

\subsection{The characterization of the ISM in LIRGs according to [FeII]1.64$\mu$m/Br${\gamma}$ ratio}

The two-dimensional distribution of the ISM in LIRGs are characterized by an observed (median) [FeII]1.64$\mu$m/Br${\gamma}$ ratio of about 0.9 that translates into an (spaxel-by-spaxel) extinction corrected ratio of about 1.5. Except for a galaxy (NGC 2369) that appears to be edge-on and heavily absorbed, the extinction-corrected 2D distributions appear as quasi-universal, i.e. has no dependence with galaxy, showing a narrow range of values with most regions clustering within factors less than 2 from the median. There are four galaxies where only for specific regions, the extinction corrected [FeII]1.64$\mu$m/Br${\gamma}$ ratio is above 5 (see Table~\ref{table:regions_fe_LIRGs}). These correspond to the nuclei of NGC 7130 (Seyfert 2) and NGC 3110, to the luminous circumnuclear [FeII] clump in NGC 5135 \citep{Colina:2012go}, and to the overall circumnuclear emission in the edge-on galaxy NGC 2369. 

Starburst models, based purely on the number of ionizing photons and the [FeII] emission associated with the expected supernova rate, predict [FeII]1.64$\mu$m/Br${\gamma}$ ratios between 0.1 and 1.4 for a continuous star formation characterized by different IMFs \citep{Colina:1993p6004}. According to these models, only star formation with relative steep IMFs (Salpeter or steeper) and low upper-mass limit ($\sim$25-30 M$_{\sun}$), would be able to produce [FeII]1.64$\mu$m/Br${\gamma}$ ratios of about 1. This situation is unlikely and other effects not taken into account in these models like supernova-induced shocks and X-ray emission in aging instantaneous bursts, or contamination by the radiation \& shocks from an AGN, could be playing a relevant role. In fact, our Seyfert galaxies have typical ratios (median \& mode) that are 2.6 times larger than those of LIRGs, indicating that the additional excitation mechanisms associated with AGNs, i.e. X-rays and shocks, could enhance the emission of [FeII] with respect to the Br$\gamma$ even in the highly ionized environments around the AGN. 

\begin{table*}[ht]
\caption{[FeII]1.64$\mu$m/Br$\gamma$ line ratio for all LIRGS. Statistical results from SINFONI/IFS measurements$^1$}
\centering
\resizebox{1.\textwidth}{!}{
{\setlength{\tabcolsep}{5pt}
\begin{tabular}{cccccccccccccc}
\hline
\hline
\noalign{\smallskip}
  Object & \multicolumn{2}{c}{Median} & \multicolumn{2}{c}{Mode} & \multicolumn{2}{c}{Mean} &\multicolumn{2}{c}{$\sigma$} & \multicolumn{2}{c}{$\rm P_{5}$} & \multicolumn{2}{c}{$\rm P_{\rm95}$} & $\rm N_{spaxel}$ \\
   & Obs & Corr & Obs & Corr & Obs & Corr & Obs & Corr & Obs & Corr & Obs & Corr &   \\
\noalign{\smallskip}
\noalign{\smallskip}
\hline
NGC2369 &  0.61 &  3.58 &  0.80 &  4.02 &  0.63 &  3.99 &  0.23 &  2.36 &  0.31 &  1.40 &  1.04 &  7.92 &  1583  \\
NGC3110 &  0.72 &  1.59 &  0.69 &  1.73 &  0.77 &  1.66 &  0.23 &  0.51 &  0.49 &  0.96 &  1.18 &  2.60 &  2719  \\
NGC3256 &  1.06 &  1.69 &  0.96 &  1.82 &  1.13 &  2.05 &  0.39 &  1.46 &  0.61 &  0.79 &  1.85 &  4.64 &  5553  \\
ESO320-G030 &  0.89 &  1.37 &  0.91 &  1.32 &  0.91 &  1.46 &  0.22 &  0.52 &  0.57 &  0.84 &  1.30 &  2.32 &  4011  \\
IRASF12115-4656 &  \nodata &  \nodata &  \nodata &  \nodata &  \nodata &  \nodata &  \nodata &  \nodata &  \nodata &  \nodata &  \nodata &  \nodata &     \nodata   \\
NGC5135 (all)&  0.95 &  1.89 &  1.07 &  1.44 &  1.21 &  2.59 &  0.99 &  2.25 &  0.49 &  1.02 &  3.24 &  7.43 &  1610   \\
NGC5135 ([SiVI])$^2$ &  1.06 &  2.29 &  1.07 &  1.96 &  1.40 &  3.03 &  1.13 &  2.40 &  0.48 &  1.06 &  4.09 &  8.32 &   898   \\
NGC5135 (no [SiVI])$^3$&  0.80 &  1.53 &  0.67 &  1.37 &  0.98 &  2.03 &  0.71 &  1.92 &  0.50 &  0.97 &  2.32 &  4.93 &   712  \\
IRASF17138-1017 &  1.05 &  2.00 &  1.14 &  2.08 &  1.07 &  2.08 &  0.36 &  0.90 &  0.53 &  0.83 &  1.68 &  3.81 &  2327  \\
IC4687 &  0.95 &  1.60 &  1.05 &  1.76 &  0.97 &  1.82 &  0.36 &  1.01 &  0.49 &  0.69 &  1.55 &  3.87 &  4212  \\
NGC7130 (all)&  1.18 &  1.96 &  0.93 &  2.01 &  1.42 &  3.27 &  0.85 &  4.12 &  0.50 &  0.75 &  3.17 & 10.36 &  1857  \\
NGC7130 ([SiVI])$^2$ &  2.08 &  4.60 &  2.03 &  4.60 &  2.20 &  6.83 &  0.87 &  6.29 &  0.89 &  1.82 &  3.86 & 21.49 &   498  \\
NGC7130 (no [SiVI])$^3$&  0.96 &  1.51 &  0.94 &  1.24 &  1.13 &  1.96 &  0.63 &  1.53 &  0.49 &  0.71 &  2.44 &  4.77 &  1359  \\
IC5179 &  0.55 &  0.82 &  0.54 &  0.85 &  0.58 &  0.94 &  0.18 &  0.63 &  0.36 &  0.42 &  0.92 &  1.72 &  4490  \\
\hline
LIRGs &  0.87 &  1.53 &  0.95 &  1.46 &  0.94 &  1.94 &  0.49 &  1.75 &  0.43 &  0.62 &  1.72 &  4.53 & 28362  \\
\hline
\hline
\end{tabular}}}
\tablefoot{$^1$ Columns indicate statistical results for the distribution of the observed ratios (obs), and of the extinction corrected values (corr). Last column (N$_{\rm spaxels}$ gives the number of spaxels for each galaxy.   $^2$ Statistics using only the spaxels were [SiVI]1.96 $\mu$m emission line has been detected. $^3$ Statistics using all the spaxels within the field-of-view with no [SiVI]1.96 $\mu$m emission line detected.}
\label{table:distrib_fe_LIRGs}
\end{table*}

\begin{table*}[ht]
\caption{H$_2$ 2.12$\mu$m/Br$\gamma$ line ratio for all LIRGS. Statistical results from SINFONI/IFS measurements$^1$}
\centering
\resizebox{1.\textwidth}{!}{
{\setlength{\tabcolsep}{5pt}
\begin{tabular}{cccccccccccccc}
\hline
\hline
\noalign{\smallskip}
  Object & \multicolumn{2}{c}{Median} & \multicolumn{2}{c}{Mode} & \multicolumn{2}{c}{Mean} &\multicolumn{2}{c}{$\sigma$} & \multicolumn{2}{c}{$\rm P_{5}$} & \multicolumn{2}{c}{$\rm P_{\rm95}$} & $\rm N_{spaxel}$ \\
   & Obs & Corr & Obs & Corr & Obs & Corr & Obs & Corr & Obs & Corr & Obs & Corr &   \\
  \noalign{\smallskip}
\noalign{\smallskip}
\hline
NGC2369 &  1.08 &  1.21 &  1.18 &  1.32 &  1.19 &  1.33 &  0.59 &  0.66 &  0.55 &  0.60 &  2.30 &  2.57 &  1583  \\
        NGC3110 &  0.80 &  0.84 &  0.80 &  0.80 &  0.85 &  0.89 &  0.33 &  0.34 &  0.43 &  0.46 &  1.46 &  1.53 &  2719  \\
        NGC3256 &  1.00 &  1.04 &  1.09 &  1.28 &  1.20 &  1.24 &  0.82 &  0.84 &  0.33 &  0.34 &  2.90 &  2.96 &  5553  \\
    ESO320-G030 &  0.74 &  0.76 &  0.69 &  0.69 &  0.99 &  1.02 &  0.84 &  0.86 &  0.35 &  0.36 &  2.46 &  2.53 &  4011  \\
IRASF12115-4656 &  0.69 &  0.73 &  0.74 &  0.74 &  0.73 &  0.77 &  0.25 &  0.26 &  0.39 &  0.41 &  1.20 &  1.25 &  3526  \\
NGC5135 (all)&  1.31 &  1.37 &  1.88 &  1.88 &  1.45 &  1.52 &  0.84 &  0.87 &  0.48 &  0.50 &  2.89 &  3.01 &  1610  \\
NGC5135 ([SiVI])$^2$&  1.46 &  1.53 &  1.99 &  1.98 &  1.54 &  1.61 &  0.76 &  0.79 &  0.55 &  0.58 &  2.82 &  2.91 &   898  \\
NGC5135 (no [SiVI])$^3$&  1.09 &  1.14 &  0.68 &  0.80 &  1.34 &  1.40 &  0.92 &  0.95 &  0.44 &  0.46 &  2.93 &  3.06 &   712  \\
IRASF17138-1017 &  0.65 &  0.68 &  0.58 &  0.93 &  0.76 &  0.79 &  0.45 &  0.47 &  0.22 &  0.23 &  1.61 &  1.68 &  2327  \\
IC4687 &  0.540 &  0.56 &  0.61 &  0.61 &  0.57 &  0.59 &  0.29 &  0.31 &  0.17 &  0.18 &  1.13 &  1.17 &  4212  \\
NGC7130 (all)&  1.05 &  1.10 &  1.11 &  1.11 &  1.42 &  1.47 &  1.24 &  1.28 &  0.26 &  0.27 &  4.20 &  4.33 &  1857   \\
NGC7130 ([SiVI])$^2$&  1.79 &  1.88 &  1.23 &  1.22 &  2.31 &  2.42 &  1.42 &  1.44 &  0.95 &  1.06 &  5.15 &  5.31 &   498  \\
NGC7130 (no [SiVI])$^3$&  0.76 &  0.79 &  0.61 &  1.11 &  1.09 &  1.12 &  0.98 &  1.01 &  0.23 &  0.24 &  3.12 &  3.21 &  1359  \\
IC5179 &  0.71 &  0.73 &  0.84 &  0.84 &  0.78 &  0.80 &  0.37 &  0.38 &  0.32 &  0.33 &  1.46 &  1.49 &  4490  \\
\hline
LIRGs &  0.77 &  0.80 &  0.79 &  0.67 &  0.97 &  1.01 &  0.73 &  0.75 &  0.29 &  0.30 &  2.30 &  2.38 & 31888  \\
\hline
\hline
\end{tabular}}}
\tablefoot{$^1$ Columns indicate statistical results for the distribution of the observed ratios (obs) and of the extinction corrected values (corr). Last column (N$_{\rm spaxels}$ )gives the number of spaxels for each galaxy. $^2$ Statistics using only the spaxels where [SiVI]1.96 $\mu$m emission line has been detected. $^3$ Statistics using all the spaxels within the field-of-view with no [SiVI]1.96 $\mu$m emission line detected.}
\label{table:distrib_h2_LIRGs}
\end{table*}

\begin{table*}[ht]
\caption{[FeII]1.64$\mu$m/Br$\gamma$ line ratio for Seyferts. Statistical results from NIFS/IFS measurements$^1$}
\centering
{
{\setlength{\tabcolsep}{5pt}
\begin{tabular}{cccccccc}
\hline
\hline
\noalign{\smallskip}
  Object & Median & Mode & Mean &$\sigma$ & $\rm P_{5}$ & $\rm P_{\rm95}$ &$\rm N_{spaxel}$ \\
  \noalign{\smallskip}
\noalign{\smallskip}
\hline
Mrk1157 &  5.43 &  4.99 &  7.05 &  5.67 &  1.43 & 18.63 &   868  \\
Mrk1066 &  3.61 &  3.49 &  3.70 &  1.16 &  1.97 &  5.81 &  1092  \\
ESO428-G014 &  4.15 &  4.62 &  4.34 &  1.21 &  2.83 &  6.91 &    60  \\
Mrk79 &  2.80 &  3.46 &  4.48 &  5.53 &  0.73 & 15.47 &  1395  \\
NGC4151 &  4.30 &  4.43 &  4.56 &  2.15 &  1.68 &  8.50 &   421  \\
\hline
Seyferts &  3.67 &  3.98 &  4.85 &  4.57 &  1.00 & 13.24 &  3836  \\
\hline
\hline
\end{tabular}}}
\tablefoot{$^1$ Derived from the original [FeII]1.26$\mu$m and Pa$\beta$ ratios without further correction for internal extinction (see $\S$2.1)}
\label{table:distrib_fe_Sy}
\end{table*}

\begin{table*}[ht]
\caption{H$_2$ 2.12$\mu$m/Br$\gamma$ line ratio for Seyferts. Statistical results from NIFS/IFS measurements}
\centering
{
{\setlength{\tabcolsep}{5pt}
\begin{tabular}{cccccccc}
\hline
\hline
\noalign{\smallskip}
  Object & Median & Mode & Mean &$\sigma$ & $\rm P_{5}$ & $\rm P_{\rm95}$ &$\rm N_{spaxel}$ \\
  \noalign{\smallskip}
\noalign{\smallskip}
\hline
Mrk1157 &  1.71 &  1.58 &  2.25 &  2.07 &  0.50 &  5.62 &   868  \\
Mrk1066 &  1.07 &  1.32 &  1.13 &  0.47 &  0.46 &  1.98 &  1092  \\
ESO428-G014 &  1.07 &  1.12 &  1.16 &  0.40 &  0.66 &  2.16 &    60  \\
Mrk79 &  1.96 &  2.04 &  2.14 &  1.10 &  0.76 &  4.37 &  1395  \\
NGC4151 &  0.52 &  0.38 &  0.78 &  0.69 &  0.23 &  1.99 &   421  \\
\hline
Seyferts &  1.41 &  1.54 &  1.71 &  1.36 &  0.40 &  4.00 &  3836  \\
\hline
\hline
\end{tabular}}}
\label{table:distrib_h2_Sy}
\end{table*}

\subsection{The characterization of the ISM in LIRGs according to the H$_2$/Br${\gamma}$ ratio}

The ISM in LIRGs is characterized by a (median) H$_2$/Br${\gamma}$ ratio of about 0.8 (see Table~\ref{table:distrib_h2_LIRGs} for details). This ratio is typically a factor $\sim$ 1.8$-$2 lower than the corresponding ratios for Seyferts (see Table~\ref{table:distrib_h2_Sy}). In addition, while Seyferts cover a wide range of  H$_2$/Br${\gamma}$ values from about 0.4 to 4.0, LIRGs share similar lower limits (0.3 for LIRGs) but have an upper limit of about 2.4 (95\% of spaxels are below this value, see Table~\ref{table:distrib_h2_LIRGs}). This suggests that the excitation mechanism is such that it transitions from LIRGs to Seyferts, i.e. from star-forming dominated galaxies (seven out of nine LIRGs) to AGN-dominated galaxies. In fact, the H$_2$/Br$\gamma$ values are consistent with those of Seyferts for the coronal [SiVI] emitting regions in the two LIRGs (NGC 5135 and NGC 1730) where this high-excitation AGN tracer has been detected (see Table~\ref{table:distrib_h2_LIRGs}). For these two galaxies, the H$_2$/Br$\gamma$ ratio in the [SiVI] emitting regions is about 1.5 - 2 times higher than in the [SiVI] non-emitting regions. 

Detailed studies of the near-IR H$_2$ line ratios in Seyferts, star-forming galaxies \citep{Riffel:2013MNRAS429}, LIRGs (\citealt{Bedregal:2009p2426}, \citealt{Emonts:2014jd}), and nearby spirals \citep{Mazzalay:2013MNRAS428} indicate that the hot molecular gas is close to thermally excited because of a combination of X-ray radiation and shocks. On the other hand, the Br$\gamma$ line is a direct tracer of the UV-ionizing radiation and therefore the H$_2$/Br$\gamma$ ratio gives an empirical first indication of the relative importance of each mechanism (UV radiation versus X-ray radiation or shocks). This will be discussed further in sections $\S$3.3.2 to $\S$3.3.4 when presenting the line ratios for the galaxies selected as prototypes.

\subsection{Understanding the near-IR emission line diagnostic diagram for star-forming galaxies and AGNs}

\subsubsection{LIRGs in the [FeII]1.64$\mu$m/Br${\gamma}$ - H$_2$ 2.12$\mu$m/Br${\gamma}$ plane}

\begin{figure*}
\centering
\includegraphics[width=12cm]{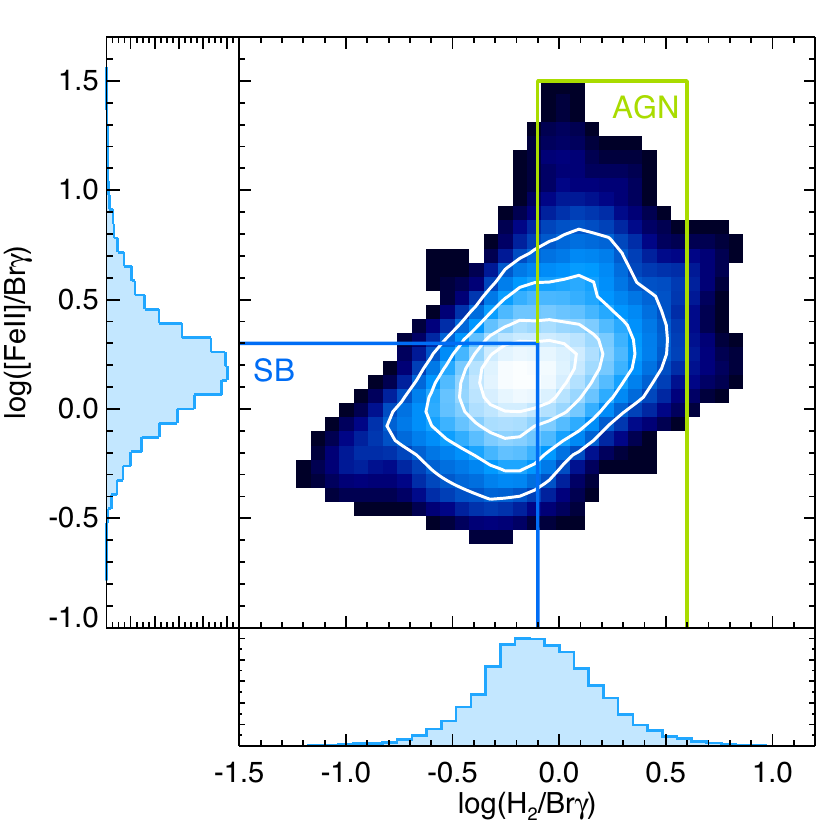}
\caption{Distribution of the spatially resolved regions (spaxel-by-spaxel) in the log([FeII]1.64$\mu$m/Br${\gamma}$) $-$ log(H$_2$2.12$\mu$m/Br${\gamma}$) plane for the sample of LIRGs. Contours encircle the areas were 25\%, 50\%, 75\%, and 90\% of the spaxels are contained. The histograms represent the distribution for the [FeII]1.64$\mu$m/Br${\gamma}$ and H$_2$/Br$\gamma$ ratios with mode (log) values of $+$0.16 and $-$0.17, respectively. The blue and green lines indicate the upper limits for the young star-forming regions and AGNs as derived from our spatially resolved two-dimensional spectroscopy on the high surface brightness compact regions of the prototype LIRGs (see section $\S$3.3 for details)}
\label{figure:LIRGs}
\end{figure*}

 The distribution of all spatially resolved regions in our sample of LIRGs ($\sim$ 28000 spaxels in nine LIRGs) covers a well-defined region in the [FeII]/Br${\gamma}$ - H$_2$/Br${\gamma}$ plane (see Figure~\ref{figure:LIRGs}), ranging from pure star-forming regions to AGNs (Seyferts), according to our empirical classification derived from the spatially resolved spectroscopy of bright AGN, compact young and aged star-forming clumps in three of our LIRGs, respectively (see sections 3.3.2 to 3.3.4).

In addition no regions  simultaneously show line ratios log(FeII]/Br${\gamma}$) $\geq$1.0 and  log(H$_2$/Br${\gamma}$) $\geq$ 0.8, i.e. LINERs according to the classification scheme proposed by Riffel and collaborators \citep{Riffel:2013MNRAS429}. These results come as a surprise as all LIRGs but two (NGC 5135 and NGC 7130) in the sample are classified as star forming and do not show evidence of a luminous AGN, according to the mid-IR \citep{AlonsoHerrero:2012p744} and X-ray emission \citep{PereiraSantaella:2011fw}. In addition, according to the optical emission line diagnostics, the excitation of the extended emission is characteristic of star formation and LINERs (\citealt{MonrealIbero:2006p2309}, \citeyear{Monreal-Ibero:2010p517}).  This extended LINER emission in LIRGs is explained by the presence of large scale shocks \citep{Monreal-Ibero:2010p517} and composite AGN-starbursts (\citealt{Kewley:2001p8174}; Yun, Kewley \& Sanders 2010), and likely not due to the existence of a weak AGN. 
In the following, we further investigate  these apparent inconsistencies by studying in more detail two LIRGs that are identified as prototypes of excitation dominated by star formation (IC 4687), by a Seyfert 2 nucleus (NGC 7130), and a third LIRG (NGC 5135) classified as composite (AGN, and star-forming clumps of different ages and masses). In this galaxy, the different excitation mechanisms (i.e. AGN and star-forming clumps of different ages) are at play.  This will help first to spatially resolve and identify the dominant ionization mechanism in each region, and secondly define new limits and relations in the near-IR plane for AGNs, pure star-forming regions (young and evolved), and composite objects. In our small sample of LIRGs, all galaxies but NGC 5135 and NGC 7130 belong to the class of star-forming galaxies with no evidence of a luminous AGN. However, it is known that a fraction of LIRGs have AGNs \citep{Veilleux:1995p98}, and therefore any given LIRG will belong to one of the categories represented by the prototypes.

Based on our capability of spatially resolving the distribution of the emitting gas with our integral field spectroscopy, we differentiate five types of regions: i) bright compact Br$\gamma$ emitting regions that are identified with young ($\leq$ 6 Myr), massive star-forming clumps dominated by the radiation from main-sequence massive stars; ii) bright compact [FeII] regions identified as aged ($\sim$ 8-40 Myr) star-forming clumps dominated by supernovae; iii) bright compact [SiVI] regions identified with nuclear AGNs; iv) diffuse extended [SiVI] emitting regions where the AGN radiation field is still detected; and v) diffuse extended non-[SiVI] emitting regions covering most of the galaxy. In the following we identify these different emitting regions in the prototypes  to more accurately constrain the line ratios produced by different ionization sources.

\subsubsection{IC 4687: prototype of LIRG dominated by star formation}

\begin{figure*}
\includegraphics[width=18cm]{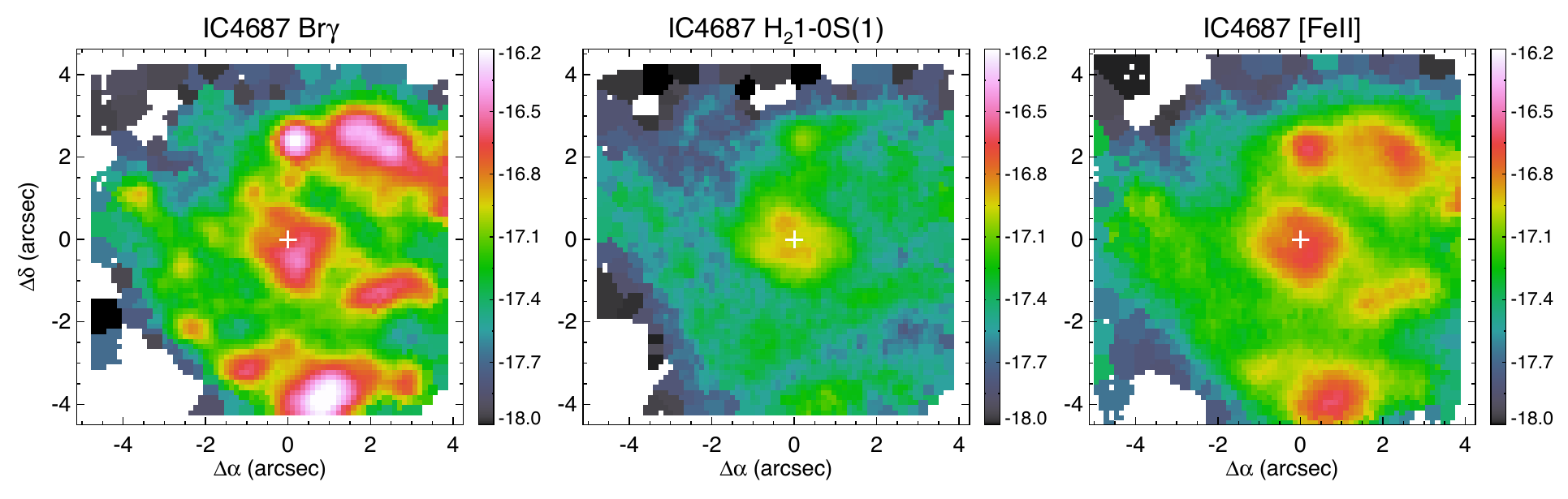}
\includegraphics[width=18cm]{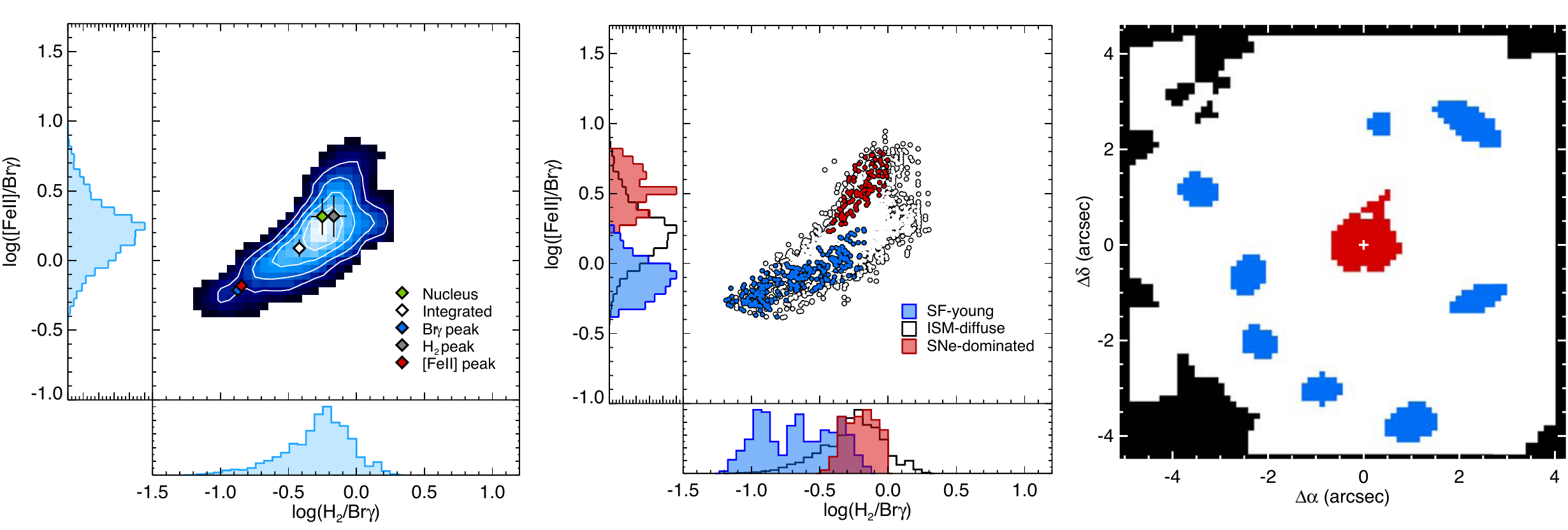}
\caption{Maps (top panels) of the ionized (Br$\gamma$, warm molecular (H$_2$ 2.12$\mu$m) and partially ionized ([FeII] 1.64$\mu$m) gas for IC 4687, prototype star-forming LIRG, obtained from the SINFONI data cubes. Based on the emission line maps, the high surface brightness regions identified as young star-forming (in blue), aged, SNe-dominated (in red), and extended diffuse (in white) are shown for reference (lower right panel). The distribution of these regions in the log([FeII]1.64$\mu$m/Br${\gamma}$) $-$ log(H$_2$ 2.12$\mu$m/Br$\gamma$) plane (lower central panel) occupy different areas indicating differences in the excitation conditions. The overall (spaxel-by-spaxel) distribution over the entire SINFONI FoV, as well as the line ratios for the nucleus and brightest emission line peaks are also shown (lower left panel).}
\label{figure:IC4687}
\end{figure*}

IC 4687 can be considered as the prototype of star-forming LIRG with its Br$\gamma$ emission (Fig.2 left upper panel) dominated by several circumnuclear star-forming clumps, and  a weak Br$\gamma$ nucleus, which is also identified as weak Pa$\alpha$ emitter on HST imaging \citep{AlonsoHerrero:2006p4703}. The H$_2$ emission (Fig.2 central upper panel) appears very diffuse and distributed over the entire circumnuclear region, with an above-average nucleus, and presenting the highest surface brightness. Finally, the [FeII] emission (Fig.2 right upper panel) follows that of the Br$\gamma$ clumpy light distribution.  The overall distribution of the line ratios in the  [FeII]1.64$\mu$m/Br${\gamma}$ - H$_2$2.12$\mu$m/Br${\gamma}$ plane (Fig.~\ref{figure:IC4687}, lower left panel) is such that the mode of the line ratios is outside the previously identified region for star-forming galaxies \citep{Riffel:2013MNRAS429} but with a long tail towards low [FeII]1.64$\mu$m/Br${\gamma}$ (0.6)  H$_2$/Br$\gamma$ ratios (down to 0.15), characteristic of star-forming regions dominated by young, ionizing stars as in M82 (\citealt{ForsterSchreiber:2001ApJ552}; see also Fig.~\ref{figure:prot}). 

The bright circumnuclear Br$\gamma$ regions (blue points/regions in Fig.~\ref{figure:IC4687}) identify young star-forming regions with [FeII]/Br$\gamma$ and H$_2$/Br$\gamma$ values well below 1. The nuclear region (red points/regions in Fig.~\ref{figure:IC4687}) appears as a weak and resolved ionizing source (as traced by the Pa$\alpha$ and Br$\gamma$ lines), with no evidence of an AGN (\citealt{AlonsoHerrero:2006p4703}, \citeyear{AlonsoHerrero:2012p744}, \citealt{PereiraSantaella:2011fw}), and is identified with evolved, supernova dominated, star-forming regions (Pereira-Santaella et al. in prep). The line ratios are well separated from that of the young star-forming regions with [FeII]/Br$\gamma$ and H$_2$/Br$\gamma$ ratios about six times higher than that of young star-forming clumps, respectively. The diffuse, line-emitting gas (white points/region in Fig.~\ref{figure:IC4687}) has [FeII]/Br$\gamma$ ratios in between those of young and supernovae dominated star-forming regions, while the H$_2$/Br$\gamma$ ratio is more consistent with that of supernovae dominated regions.

\subsubsection{NGC 7130: prototype of a LIRG dominated by a Seyfert 2 nucleus}

\begin{figure*}
\includegraphics[width=17cm]{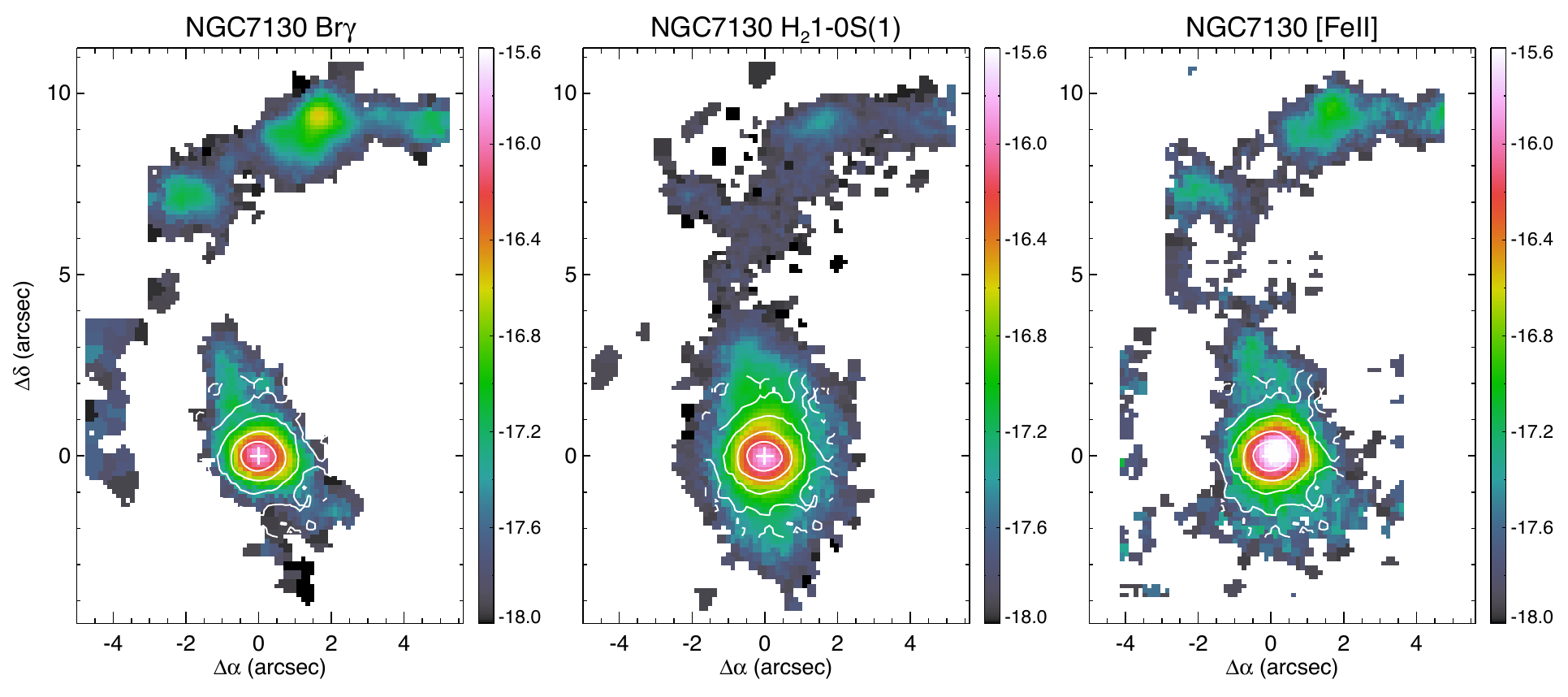}
\includegraphics[width=17cm]{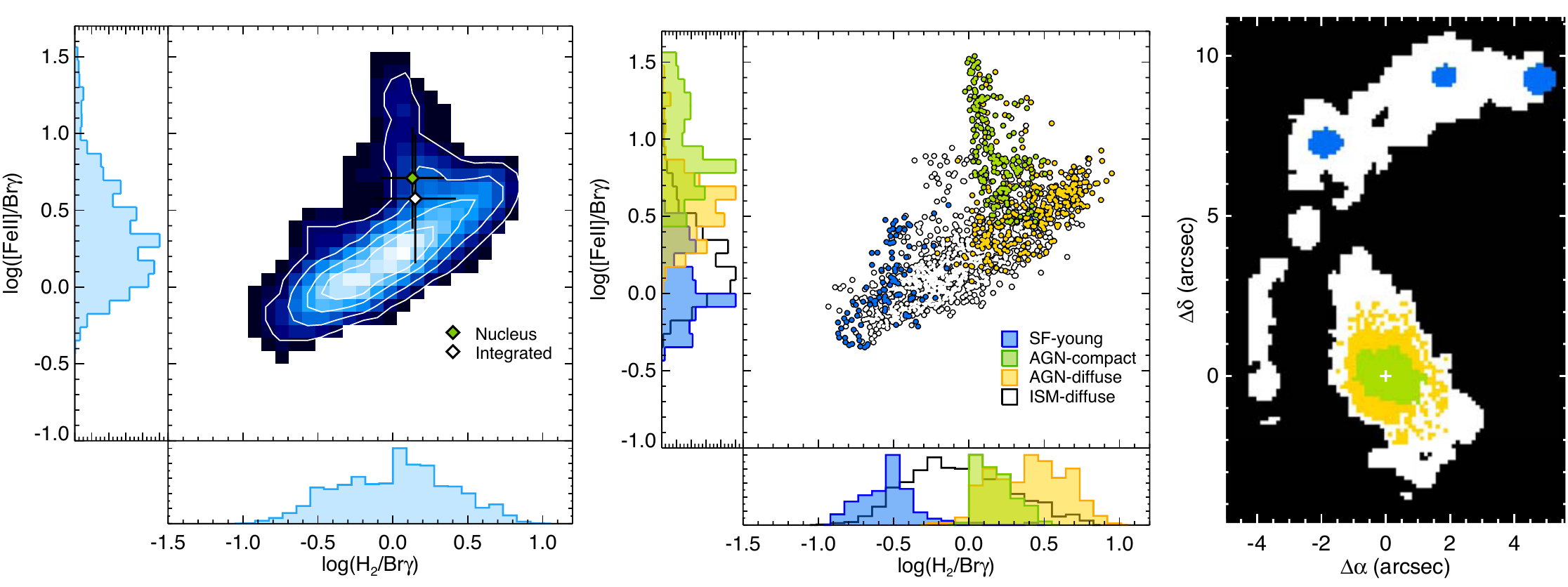}
\caption{Same as Fig.~\ref{figure:IC4687}, except for the AGN prototype LIRG NGC 7130. Note that the coronal [SiVI]1.96$\mu$m emission line is detected in the spectra of this galaxy (contours in top panels) and therefore two new regions, AGN compact (green points) and AGN diffuse (yellow points) in the nucleus, and around it have been identified and occupy a region in the log([FeII]1.64$\mu$m/Br${\gamma}$) $-$ log(H$_2$ 2.12$\mu$m/Br$\gamma$) plane very different from that of young star-forming clumps. Also note that for NGC 7130, the nucleus coincides with the peak emission for all lines.}
\label{figure:NGC7130}
\end{figure*}

The light distribution of all emission lines in this galaxy is dominated by the nucleus and its surrounding regions while a few star-forming clumps appear along a spiral arm at distances of about 2.4 to 3 kpc from the nucleus (see Fig.~\ref{figure:NGC7130}). The nucleus itself contains a known Seyfert 2-type AGN (\citealt{Levenson:2005ApJ618}, \citealt{AlonsoHerrero:2012p744}), also identified in our spectra by the presence of coronal [SiVI]1.96$\mu$m emission line \citep{Piqueras2012A&A546A}. In this galaxy, there are two well-identified distributions in the near-IR diagnostic plane (Fig.3 lower left panel). While a large fraction of the spatially resolved regions are distributed  between the upper right corner of the star-forming region and the low AGN line ratios, there is a non-negligible fraction of spaxels showing very high [FeII]/Br$\gamma$ values ($\geq$ 5), as well as H$_2$/Br$\gamma$ values similar to those of Seyferts (i.e. $\sim$1-2). Note that previous optical and ultraviolet studies of the nucleus of this galaxy have identified it as an AGN plus star-forming region. Optical emission line ratios  show a smooth transition from AGN-dominated in the centre to star-forming activity in the external regions (\citealt{Davies:2014co} and references), while the observed UV and mid-IR spectrum of the nucleus indicates the presence of young star clusters \citep{GonzalezDelgado:1998ApJ505} and   PAH emission characteristic of star formation (\citealt{DiazSantos:2010ApJ723}, \citealt{Esquej:2014ApJ780}), respectively.

In this galaxy, we identify four regions (Fig.3 bottom panels): i) the compact [SiVI] emitting region  (green points/region) as the AGN-dominated nucleus; ii) the compact and bright Br$\gamma$ regions (blue points/regions) as the young star-forming clumps in the spiral arm; and iii) the diffuse AGN ionized region (yellow points/region) where weak [SiVI] emission around the nucleus is still detected; and finally iv) the diffuse ionized interstellar medium (white points/regions) in the extranuclear regions where no [SiVI] is detected, and along the spiral arm. 

The AGN-dominated nucleus clearly separates from the bright, compact Br$\gamma$ regions and from the diffuse ionized medium, covering exclusively the highest (i.e. $\geq$ 5) [FeII]/Br$\gamma$ values. On the other hand, the bright, Br$\gamma$ circumnuclear regions, cover a range of values similar to that of the regions dominated by young stars in the prototype of star-forming LIRG IC 4687 ($\S$3.3.2). The AGN-diffuse medium clearly separates from the AGN- and young-star-dominated regions, covering the highest end of the H$_2$/Br$\gamma$ ratio while presenting intermediate [FeII]/Br$\gamma$ values. Lastly, the diffuse non-[SiVI] emitting medium covers a very wide range of H$_2$/Br$\gamma$ values from those of compact young star-forming clumps to AGN-dominated nucleus, and with  [FeII]/Br$\gamma$ values slightly above that for young star-forming regions.

\subsubsection{NGC 5135: prototype of composite LIRG}

\begin{figure*}
\centering
\includegraphics[width=17cm]{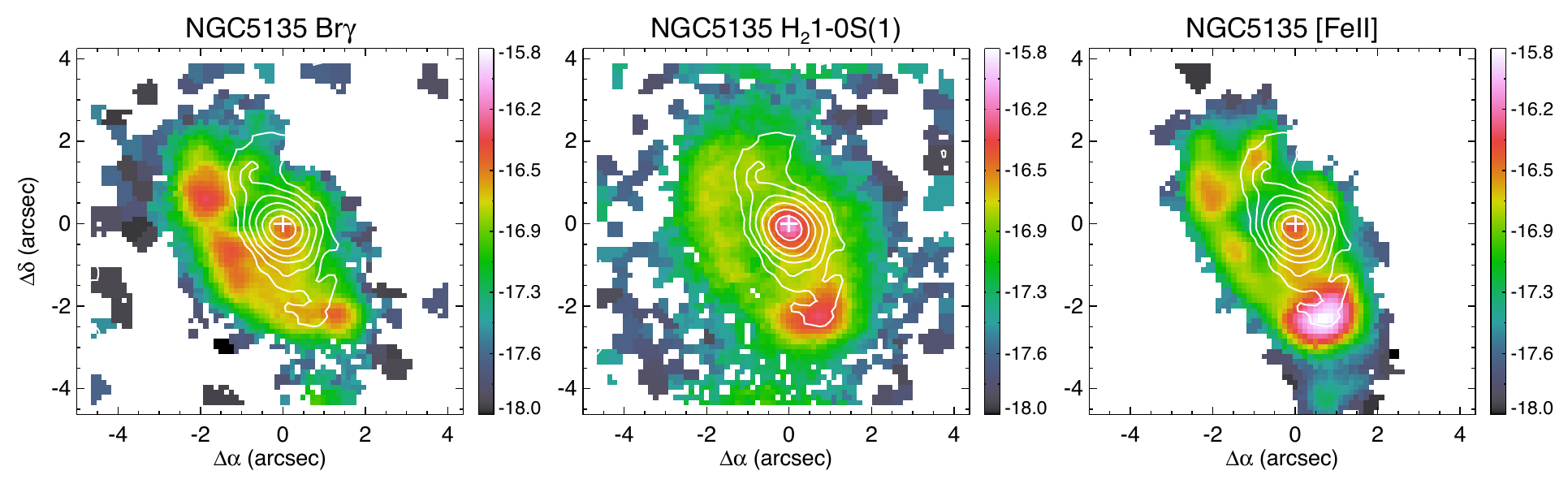}
\includegraphics[width=17cm]{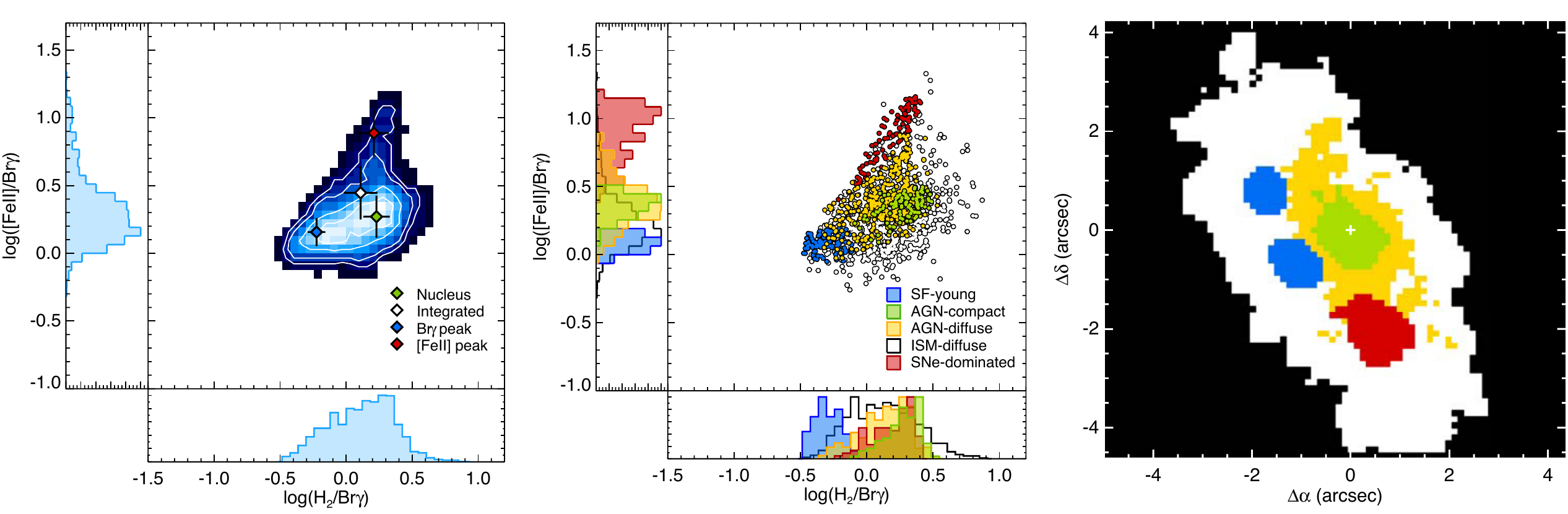}
\caption{Same as Fig.~\ref{figure:IC4687}, except for the composite prototype LIRG NGC 5135. This galaxy is the only example where all different types of regions are identified: circumnuclear young, and SNE-dominated, star-forming clumps, compact and extended AGN regions (traced by the [SiVI]1.96$\mu$m emission line and shown by contours in top panels), and general diffuse ISM. All these different regions, i.e. excitation mechanisms, occupy different and well-separated areas in the log([FeII]1.64$\mu$m/Br${\gamma}$) $-$ log(H$_2$ 2.12$\mu$m/Br$\gamma$) plane (lower midle panel).}
\label{figure:NGC5135}
\end{figure*}

Out of our sample of LIRGs, NGC 5135 is the only galaxy where clear evidence for spatially resolved and well-separated regions, dominated by either a Seyfert 2 nucleus, young massive stars, or supernovae, is well established (\citealt{Bedregal:2009p2426}; \citealt{Colina:2012go}). Imaging of the different emission lines for the entire emitting region of about 1.5 kpc across, for the regions dominated by the AGN, young stars, supernovae, and diffuse ISM, and the specific integrated values for these regions are presented in Fig.~\ref{figure:NGC5135}. All the regions are within a radius of 600 parsecs from the nucleus, and are traced by the brightest  circumnuclear Br$\gamma$ (i.e. young star-forming clumps), circumnulear [FeII]  (i.e. aged, SNe dominated star-forming clump), and compact [SiVI] (i.e. AGN, shown as contours in Fig. 4 upper panels) emitting regions. In addition, the diffuse surrounding medium is traced by the lower surface brightness areas outside these compact regions. As for NGC 7130, we can distinguish the diffuse AGN-contaminated medium traced by the presence of extended, low surface brightness [SiVI], and the general diffuse ionized medium where no trace of coronal gas is detected. NGC 5135 represents therefore a prototype of the so-called composite galaxies where AGN, young stars, and supernovae clumps, as well as AGN-contaminated and general diffuse ionized medium, are all within about 1-2 kpc, therefore unresolved (even with AO) in intermediate and high-redshift galaxies. As for IC4687 and NGC 7130, the comparison of the distributions of the emission line ratios for different, spatially resolved regions (Fig. 4) give very important clues:

\textit{i}) The young ($\leq$ 6 Myr) star-forming clumps dominated by massive, ionizing stars (blue points/regions) appear to have [FeII]/Br$\gamma$ and H$_2$/Br$\gamma$ ratios factors of about six and three lower than that of the aged, SNe-dominated, star-forming clump (red points/region), respectively. These young clumps occupied a region similar to that of the young stars dominated IC4687 galaxy and the star-forming clumps in the NGC 7130 spiral. 

\textit{ii}) The aged ($\sim$ 8-40 Myr), SNe-dominated circumnuclear star-forming clump (red points/region) occupies a region with very high [FeII]/Br$\gamma$ ratios ($\sim$ 3 to 16). The lower range is similar to the ratios measured in the brightest radio supernovae remnants of M82 \citep{Greenhouse:1997ApJ476}. The upper values are close to those detected in supernovae remnants, such as the Crab nebula \citep{Graham:1990ApJ352} and IC 443 \citep{Graham:1987ApJ313}, where FeII]/Br$\gamma$ ratios well above 10 and up to 50 have been measured. 

\textit{iii}) The AGN-dominated nuclear region (green points/region) has an H$_2$/Br$\gamma$ ratio similar to that of the SNe-dominated clump while having a [FeII]/Br$\gamma$ ratio $\sim$ 3.5 lower than that of the SNe-dominated region, i.e. close to a factor 2 higher than that of the young star-forming clumps. 

\textit{iv}) The AGN diffuse region (yellow points/region) traced by the low surface brightness [SiVI] emitting gas occupies mostly the region of AGN-dominated ionized gas, with some overlap with the SNe-dominated and well disentangled from the young star-forming clumps.

\textit{v}) The diffuse ionized interstellar medium in the external regions (white points/region) has an [FeII]/Br$\gamma$ ratio distribution in between that of AGN and young star-forming clumps, and an H$_2$/Br$\gamma$ distribution wider than any of the other regions but covering mostly the range of values characteristic of AGN- and SNe-dominated star-forming clumps.

All these characteristics can be understood as variations on the properties of the radiation field, presence of shocks, and iron abundances on scales of hundred of parsecs or less. The differences between the aged ($\sim$ 8-40 Myr) SNe-dominated and the young  ($\leq$ 6 Myr) star-forming clumps can be understood as the combination of three different effects: \textit{i}) enhancement of the iron abundance and emission due to supernovae (i.e. increased [FeII]); \textit{\textit{i}i}) a larger relative amount of ionizing photons in young clumps (i.e. increased Br$\gamma$); and \textit{iii}) a larger relative importance of shocks and X-ray emission associated with supernovae (i.e. increased H$_2$). On the other hand, the differences between the AGN-dominated region and the young star-forming clumps can be understood as due to a larger amount of X-rays with respect to UV-ionizing radiation in the AGN than in young, massive star-forming clumps. An excess of X-rays penetrate more deeply in the surrounding medium, producing a larger partially ionized region where [FeII] and H$\_2$ emission is enhanced.  Detailed studies of the X-ray emission in NGC 5135 \citep{Colina:2012go} established the presence of hard X-ray emitting regions associated with the Seyfert 2 nucleus and the bright SNe-dominated [FeII] emission peak, with a diffuse soft X-ray component covering the entire circumnuclear region. Moreover, the K-band H$_2$ emission line ratios for the nucleus and the SNe-dominated region are consistent with the predicted thermal ratios due to X-rays heating, while the ratios for  young star-forming clumps depart from thermal \citep{Bedregal:2009p2426}. 

The excitation properties of the diffuse gas, on the other hand, can be understood qualitatively as a consequence of the leaking of different radiation fields associated with different, spatially distributed ionization sources. In fact, the H$_2$/Br$\gamma$ ratio appears to be more likely associated with the leaking of X-rays from the AGN and SNe-dominated clump, while the [FeII]/Br$\gamma$ seems to be more associated with the UV-radiation field, assuming the same iron abundances for the nucleus, young star-forming clumps and diffuse medium. While there is no evidence for radio jets in this galaxy, shocks associated with jets can also enhance the [FeII] emission as in NGC 4151 \citep{StorchiBergmann:2009if}, and therefore this mechanism should also be considered in AGNs with energetic radio jets. 

\section{Discussion}

\subsection{LIRGs and Seyferts. Defining new near-IR line ratio limits and relations to discriminate activity in galaxies}

\begin{figure*}
\centering
\includegraphics[width=15cm]{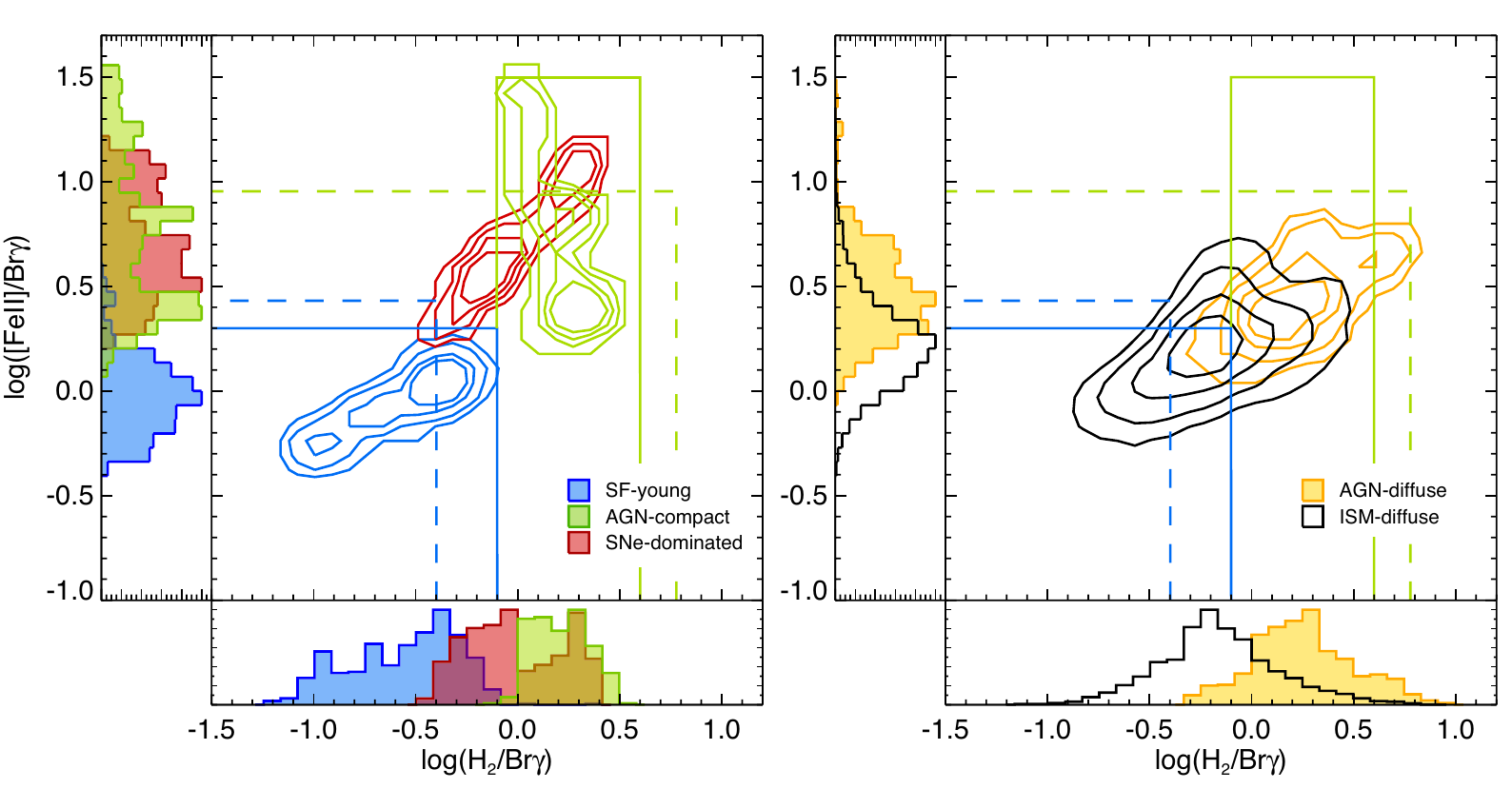}
\caption{Distribution of the line ratios in the log(FeII]1.64$\mu$m/Br$\gamma$) $-$ log(H$_2$2.12$\mu$m/Br$\gamma$) plane for the various types of regions identified in the three prototypes (IC4687, NGC 7130, and NGC 5135). Left panel: compact high surface brightness young star-forming clumps (blue), aged, SNe-dominated clumps (red), compact nuclear AGN (green). Right panel: diffuse AGN (yellow), and general diffuse medium (white). The blue and green lines indicate the new upper limits for the young star-forming regions and Seyferts as derived from our spatially resolved two-dimensional spectroscopy on the high surface brightness compact regions. For comparison, the two broken lines represent the upper limits of the line ratios for the starbursts and Seyferts  previously identified by Riffel and coworkers \citep{Riffel:2013MNRAS429} based on 1D spectroscopy. Contours contain 25\%, 50\%, 75\%, and 90\% of the points for each class.}
\label{figure:prot}
\end{figure*}

\begin{figure*}
\centering
\includegraphics[width=18cm]{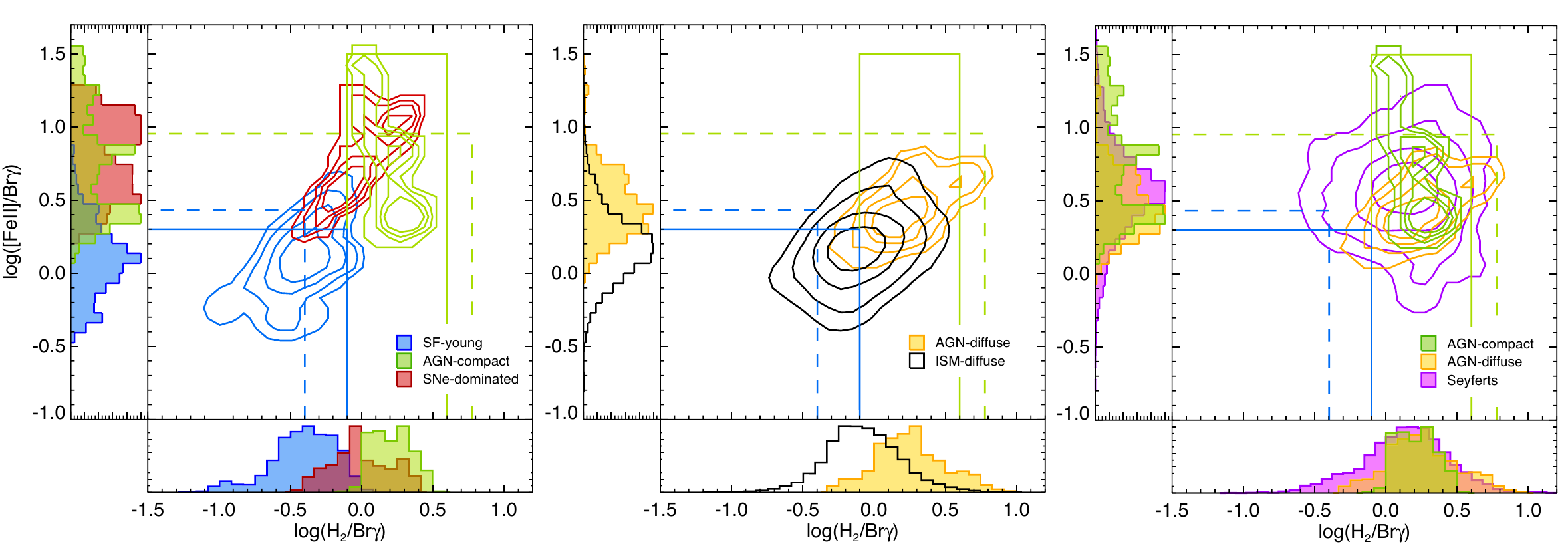}
\caption{Distribution of the line ratios for all LIRGs in the log(FeII]1.64$\mu$m/Br$\gamma$) $-$ log(H$_2$2.12$\mu$m/Br$\gamma$) plane for different types of regions: high surface brightness clumps and nucleus (left panel), for the diffuse, AGN and non-AGN, extended regions (central panel), and for the compact and diffuse AGN medium in LIRGs and low-z Seyferts (right panel). The blue and green lines indicate  new upper limits for young star-forming regions and Seyferts as derived from the compact, high surface brightness regions in our LIRG prototypes (see Fig.~\ref{figure:prot}). For comparison, the two broken lines represent the limits for the starbursts and Seyferts previously identified by Riffel and coworkers \citep{Riffel:2013MNRAS429} based on 1D spectroscopy. Contours contain 25\%, 50\%, 75\%, and 90\% of the points for each class.}
\label{figure:all}%
\end{figure*}

Based on long-slit, near-IR spectra of a sample of star-forming galaxies and LINERs combined with previous similar data for Seyferts (\citealt{Rodriguez-Ardila:2005p364}, \citealt{Riffel:2013MNRAS429}) identified some areas in the [FeII]1.26$\mu$m/Pa$\beta$ and H$_2$2.12$\mu$m/Br$\gamma$ plane that discriminate among the different classes of nuclear activity in galaxies from star formation to Seyferts and LINERs. Following this study, we have transformed the original [FeII]1.26$\mu$m/Pa${\beta}$ limits to [FeII]1.64$\mu$m/Br${\gamma}$ applying the corresponding atomic and recombination parameters and correcting for internal extinction as explained in section $\S$2. These regions, identified with broken lines in Figs.~\ref{figure:prot} \& \ref{figure:all}, are defined as
\textit{i}) star-forming galaxies characterized by log([FeII]1.64$\mu$m/Br$\gamma$) $\leq +0.42$ and  log(H$_2$2.12$\mu$m/Br$\gamma$) $\leq -0.4$;
\textit{ii}) Seyfert galaxies characterized by +0.42 $<$ log([FeII]1.64$\mu$m/Br$\gamma$) $\leq +0.95$ and $-$0.4 $<$ log(H$_2$2.12$\mu$m/Br$\gamma$) $\leq +0.78$; and \textit{iii}) LINERs characterized by log([FeII]1.64$\mu$m/Br$\gamma$) $> +0.95$ and log(H$_2$2.12$\mu$m/Br$\gamma$) $> +0.78$.

While previous work (\citealt{Riffel:2013MNRAS429} and references therein) gave the flux-weighted emission line ratios and corresponding limits for Seyferts, starbursts, and LINERs as above, we  identify new limits and relations for the different excitation mechanism as a function of location within the galaxy (Fig.~\ref{figure:prot} for the prototypes, and Fig.~\ref{figure:all} for all LIRGs in the sample).  These include those either associated with the compact AGN (green contours), (circum)nuclear young and aged, SNe-dominated star-forming clumps (blue and red contours, respectively), the extended AGN-diffuse medium around the nucleus (yellow contours), and the general diffuse interstellar medium, not directly associated with any of the previous regions (black contours). This new approach, 2D distribution versus integrated light, is important because many active galaxies, including high-redshift galaxies, do actually have an AGN and (circum)nuclear star-forming clumps of different ages. This coexistence  produces  a mixture of ionizing and excitation conditions in the ISM due to the relative contribution of the different radiation fields, to the presence of shocks induced by the AGN, radio jets, stellar winds and supernovae, and to the spatial distribution of the ionizing sources.

We identify the new areas in the [FeII]1.64$\mu$m/Br$\gamma$ $-$ H$_2$2.12$\mu$m/Br$\gamma$ plane (Fig.~\ref{figure:prot}) occupied by the spatially separated and well-resolved high surface brightness regions that we associate with different dominant ionization/excitation sources in our prototypes (see $\S$3.3.2 to 3.3.4):

\begin{itemize}

\item{SF-young (i.e. $\leq$ 6 Myr, star-forming clumps): log([FeII]/Br$\gamma$)= 0.238 + 0.476 $\times$ log(H$_2$/Br$\gamma$) in the region limited by $-0.4 \leq$ log([FeII]/Br$\gamma$) $\leq +0.4$  and $-1.2 \leq$ log(H$_2$/Br$\gamma$) $\leq$ $-$0.1}

\item{SNe-dominated (i.e. $\sim$ 8-40 Myr, star-forming clumps): log([FeII]/Br$\gamma$)= 0.705 + 1.000 $\times$ log(H$_2$/Br$\gamma$) in the region limited by $+0.2 \leq$ log([FeII]/Br$\gamma$) $\leq +1.2$  and  $-0.4 \leq$ log(H$_2$/Br$\gamma$) $\leq$ +0.4}

\item{AGN-compact: log([FeII]/Br$\gamma$)= 1.009 $-$ 1.312 $\times$ log(H$_2$/Br$\gamma$) within the region limited by $-0.3 \leq$ log([FeII]/Br$\gamma$) $\leq +1.5$  and  $-0.3 \leq$ log(H$_2$/Br$\gamma$) $\leq$ +0.9}

\end{itemize}

In addition to the classification based on the compact, high surface brightness regions, limits and relations for the diffuse medium can also be established. The AGN-diffuse region identified in NGC 5135 and NGC 7130 (yellow points in Figs.~\ref{figure:prot} \& Fig.~\ref{figure:all}) are all within the AGN area as defined by \cite{Riffel:2013MNRAS429}, cover an area similar to that of the AGN-compact regions identified above but with  high values  that are less extreme in [FeII]/Br$\gamma$ (no values above 10 are detected), and are characterized by a linear relation log([FeII]/Br$\gamma$)= 0.353 + 0.458 $\times$ log(H$_2$/Br$\gamma$). The slope of this relation is very close to that found for the young star-forming clumps but clearly departs from that derived for a sample of star-forming galaxies, Seyferts, and LINERs (log([FeII]/Br$\gamma$)= 0.446 + 0.749 $\times$ log(H$_2$/Br$\gamma$), obtained from the integrated [FeII]1.26$\mu$m/Pa$\beta$ and H$_2$/Br$\gamma$ ratios \citep{Riffel:2013MNRAS429}. A possible cause for this discrepancy could be produced by the galaxies identified as LINERs in Riffel\'s sample that appear showing the highest [FeII]/Br$\gamma$ and H$_2$/Br$\gamma$ line ratios. LINERs are known to have very weak emission lines in the near-IR and therefore the subtraction of the stellar (hydrogen) absorption lines and continuum is essential to obtain an accurate measurement of the emission line fluxes as already pointed out previously (\citealt{Larkin:1998p5053}, \citealt{AlonsoHerrero:2000ApJ530}, \citealt{Riffel:2013MNRAS429}). The net effect would be an increase in the flux of the Pa$\beta$ and Br$\gamma$ lines therefore bringing down the corresponding line ratios, and therefore decreasing the slope of the linear relation. The regions identified as AGN compact and AGN diffuse agree very well with the distribution observed in Seyferts, with the mode values of their distributions in the near-IR diagnostic plane in agreement within 0.2dex (see Fig. 6 right panel). 

Finally, the diffuse interstellar medium (black contours in Fig. 6) has a very wide range of values covering those corresponding to all different excitation mechanisms. This suggests that the general heating and radiation field in these extended regions is a mixture of that of the individual heating sources (i.e. young stars, supernovae, AGN, shocks, etc.) weighted by their relative flux contribution and spatial distribution. 

This new classification scheme allows us not only to discriminate empirically the excitation/ionization dominated by young stars from that dominated by AGN but also adds a new class, the SNe-dominated (i.e. aged) star-forming regions. In addition it establishes that the excitation/ionization conditions of the extended, diffuse medium is a complex mixture of all the different mechanisms at work, weighted by their relative contribution to the total energy output and also weighted by their spatial distribution.

\subsection{Excitation mechanisms: bright, compact line-emitting regions}

\begin{figure*}
\centering
\includegraphics[width=17cm]{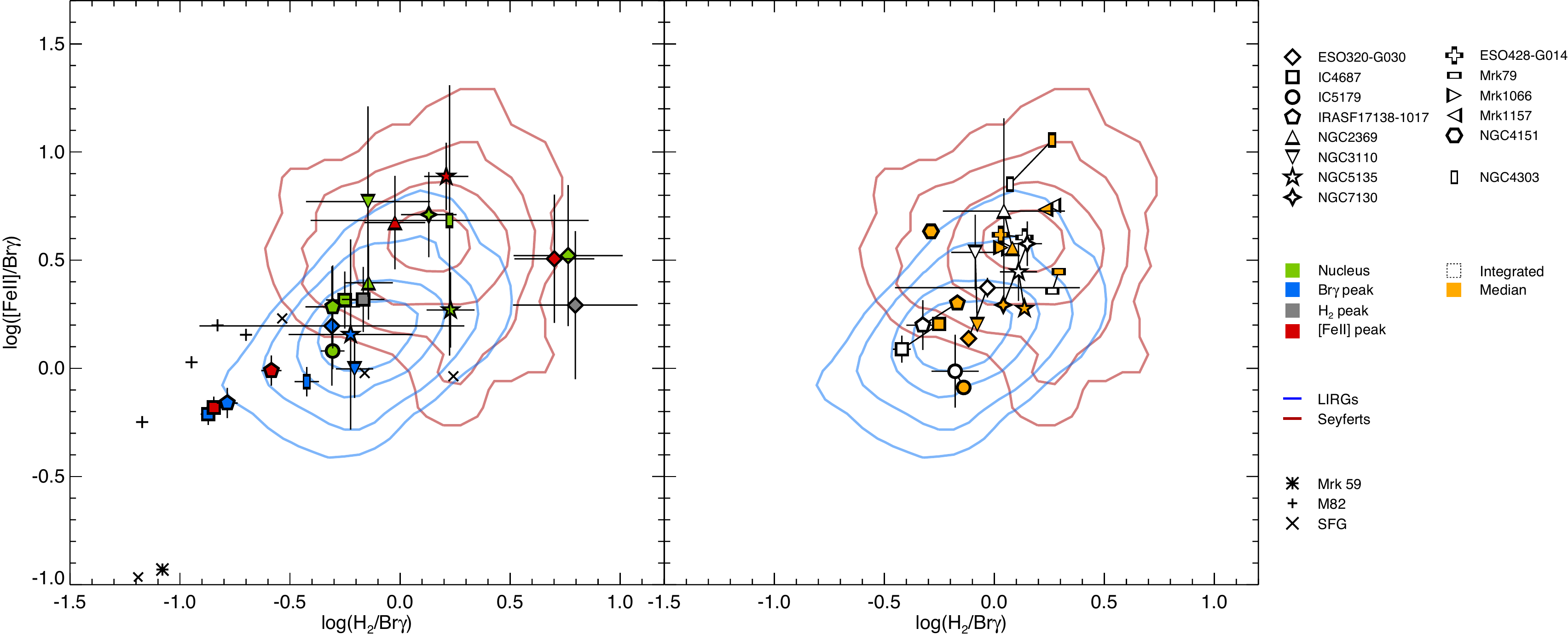}
\caption{Near-IR emission line ratios for the nucleus  and brightest Br$\gamma$, H$_2$, and [FeII] emitting regions in LIRGs (left panel). Note that in some LIRGs these regions coincide (see Tables~\ref{table:regions_fe_LIRGs} \& \ref{table:regions_h2_LIRGs} for details). The distribution of spatially resolved regions (spaxel-by-spaxel) are represented by contours for both LIRGs (blue) and Seyferts (red). Contours for LIRGs and Seyferts contain 25\%, 50\%, 75\%, and 90\% of the points. Examples of nearby star-forming galaxies like M82, Mrk 59, composite AGN+stars (NGC 4303), and other low-z SFG are also indicated. The emission line ratios for the sample of LIRGs and Seyferts as derived from the integrated values (i.e. 1D flux-weighted), and median of the 2D distribution (i.e. spaxel-by-spaxel) of values is also presented (right panel).}
\label{figure:points}
\end{figure*}

The line ratios for the nucleus, and the brightest Br$\gamma$, [FeII], and H$_2$ regions for each LIRG in the sample have also been computed (Fig.~\ref{figure:points} left) and compared with the 2D distributions for LIRGs and Seyferts (blue and red contours, respectively). Note that for some of the galaxies, the regions emitting the brightest lines are spatially coincident within the available angular resolution. In these cases, only one of the regions is represented, typically the nucleus, which is also in most cases the H$_2$ peak region (see Tables~\ref{table:regions_fe_LIRGs} \& \ref{table:regions_h2_LIRGs} for the specific values and details about the spatial coincidence of the emitting regions). In addition,  line ratios for the nucleus and star-forming regions in nearby galaxies covering different metallicities and morphologies (see $\S$2.3 for details) are also included in this analysis for a direct comparison.

Most of the brightest emitting regions are located either close to the mode of the 2D distribution of the emission line ratios for LIRGs (blue contours), or towards lower [FeII]/Br$\gamma$ and H$_2$/Br$\gamma$ ratios. These agree with the range of values observed in the nucleus and off-nuclear regions of M82, in the circumnuclear young star cluster in NGC 4303, and in the nucleus of star-forming galaxies with metallicities around solar, and are located well above the values derived in low-metallicity (Z$_{\odot}$/3 or less), star-forming galaxies (Mrk 59 and NGC 1569W). A small fraction of regions are located in the area previously identified as SNe-dominated (nucleus of NGC 3110, [FeII] peaks in NGC5135 and IC 4687) and AGN-dominated (NGC 7130).  The two galaxies with Seyfert 2 nucleus show a very different behaviour. While the Seyfert 2 nucleus of NGC 7130 lies close to the locus of the nearby Seyferts (red contours) and close to the nucleus of the  NGC 4303 (low-luminosity AGN plus nuclear young cluster, see references in $\S$2.3), the nucleus of NGC 5135 shows a significantly lower [FeII]/Br$\gamma$ ratio. In addition to an AGN, NGC 7130, and NGC 5135 have strong starbursts in their central 100 pc region (\citealt{GonzalezDelgado:1998ApJ505}, \citealt{DiazSantos:2010ApJ723}, \citealt{Esquej:2014ApJ780}). Therefore changes in the emission line ratios could be expected if the nuclear radiation field is a mixture of AGN and young stars with different relative contributions, and/or due to aging effects in the nuclear star clusters. 
Finally, only the nuclear regions in ESO320$-$G030 appear as outliers with very high values ($\sim$ 6) for the H$_2$/Br$\gamma$ ratio, and intermediate [FeII]/Br$\gamma$ (Fig.~\ref{figure:points}, see also Fig.~\ref{figure:ESO320}). We identified this region as H$_2$-dominated requiring  detailed investigation.

In summary, most compact bright regions in LIRGs appear to be similar to solar-metallicity, star-forming clumps in nearby galaxies, and well separated from low-metallicity, star-forming clumps. A smaller fraction of regions appear to be consistent with Seyfert nuclei, although the contribution of young stars and SNe appears to play a relevant role. The nucleus of ESO320$-$G030 is identified as an outlier to the general behaviour because of  its strong H$_2$ emission relative to that of Br$\gamma$.

\subsection{Spatially resolved versus integrated line ratios. Implications for the classification of high-z star-forming galaxies}

While all the analysis we performed so far is based on the two-dimensional distribution of the line ratios, most studies of the excitation mechanisms in large galaxy surveys have been based on the nuclear/integrated line ratios. The main differences between these two measurements is that, while the integrated values represent a flux-weighted measurement and therefore it is dominated by the brightest observed regions, the median of the distribution of the spatially resolved regions gives a value more representative of the general excitation and ionization conditions of the extended diffuse medium in galaxies. This is particularly important for LIRGs, where different ionizing sources coexists, and where the clumpy structure of the dust extinction can play a relevant role. 

A comparison of the flux-weighted (i.e. integrated) and the 2D distribution (median) values for the LIRGs and Seyferts is presented (Figs.~\ref{figure:points}, right panel). For the Seyferts in the sample, the integrated and median values for the line ratios agree within less than 0.1dex, and therefore occupy the same location in the [FeII]/Br$\gamma$ $-$ H$_2$/Br$\gamma$ plane, near the locus of the 2D distribution. On the other hand, clear differences appear for LIRGs.  While IC5179 shows integrated and median values within 0.1dex, IC4687 and IRASF17138$-$1017 show integrated values slightly lower (factors 0.5 to 0.6) than their median ratios. The rest of the LIRGs show larger differences in the [FeII]/Br$\gamma$ ratio with integrated over median values $\sim$ 1.5 $-$ 2.2 times higher. The net effect for these galaxies is that they appear with higher excitation, i.e. moving towards the locus of the AGNs, when the flux-weighted integrated values are considered instead of the median. Whether this is a general behaviour in LIRGs or galaxies in general (similar changes in the line ratios when comparing nuclear and off-nuclear regions in star-forming galaxies have been reported recently by \citealt{Martins:2013gq}), needs further exploration with a larger volume of galaxies with available integral field spectroscopy.  This would have obvious implications when classifying high-z galaxies for which the linear resolution would be in general very limited ($\sim$ 1.5$-$2 kpc), even with sub-arcsec angular resolution with current and future facilities like the JWST.

\section{Conclusions}

We investigated the two-dimensional excitation structure of the interstellar medium in a sample of low-z LIRGs and Seyferts, using near-IR integral field spectroscopy. Based on the spatially resolved spectroscopy of three (selected) LIRG prototypes, we identified in the [FeII] 1.64$\mu$m/Br$\gamma$ $-$ H$_2$2.12$\mu$m/Br$\gamma$ plane the regions dominated by the different heating sources, including young main-sequence massive stars, supernovae, and AGNs. This extends to the near infrared the well-known optical emission line diagnostics used to classify the activity in galaxies. The main conclusions are:

\begin{enumerate}
\item The two-dimensional (i.e. spaxel-by-spaxel) distribution of the excitation conditions of the ISM in LIRGs occupy a wide region in the near-IR diagnostic plane from $-$0.6 to $+$1.5 and from $-$1.2 to $+$0.8 (in log units) for the [FeII]/Br$\gamma$ and H$_2$/Br$\gamma$ line ratios, respectively. The corresponding median(mode) ratios are +0.18(0.16) and +0.02($-$0.04), respectively.

\item The two-dimensional distributions of the line ratios of LIRGs and nearby Seyferts partially overlap, with Seyferts showing on average larger values by factors $\sim$ 2.5 and $\sim$1.7 for the [FeII]/Br$\gamma$ (4.9 vs. 1.9) and H$_2$/Br$\gamma$ (1.7 vs 1.0) ratios, respectively.

\item We identified a range of line ratios for the different ionization/heating sources  thanks to our integral field spectroscopy of well-resolved and spatially separated, high surface brightness, line emitting clumps: \textit{i}) AGN dominated, \textit{ii}) young stars dominated, and \textit{\textit{ii}i}) supernovae dominated. These three heating mechanisms occupy mostly well-separated regions in the log([FeII]/Br$\gamma$) $-$ log(H$_2$/Br$\gamma$) plane. 

\item New areas in the near-IR diagnostic emission line plane are defined for  different heating sources with the following limits in the log([FeII]/Br$\gamma$) and log(H$_2$/Br$\gamma$) ratios, respectively: AGN-dominated: [$-$0.3, +1.5] and [$-$0.3, +0.9]; young stars-dominated: [$-$0.4, +0.4] and  [$-$1.2, $-$0.1]; and SNe-dominated: [$+$0.2, $+$1.2]  and  [$-$0.4, +0.4]. Within these limits, the 2D distribution of the regions dominated by each of the mechanisms appears to cluster around linear relations given by: i) AGN-dominated: log([FeII]/Br$\gamma$)= 1.009 $-$ 1.312 $\times$ log(H$_2$/Br$\gamma$); ii) young stars-dominated: log([FeII]/Br$\gamma$)= 0.238 + 0.476 $\times$ log(H$_2$/Br$\gamma$); and iii) SNe-dominated: log([FeII]/Br$\gamma$)= 0.705 + 1.000 $\times$ log(H$_2$/Br$\gamma$). 

\item We also identified the excitation conditions of the diffuse, extended ISM outside the bright clumps. The diffuse ISM occupies a large region that completely overlaps with that of the individual excitation mechanisms (i.e. AGN, young stars, and supernovae), indicating that its radiation/heating field is due to a mixture of the different ionization sources, weighted by their relative flux contribution and  spatial distribution.  

\item Compared with the median of the two-dimensional distribution, the integrated line ratios in LIRGs appear compatible with higher excitation conditions, i.e. towards AGNs. If this behaviour is general, it would have clear consequences when classifying high-z, star-forming galaxies based on their near-infrared integrated spectra. A similar situation could also happen with the optical emission line ratios that would deserve detailed exploration with current and future observatories like the James Webb Space Telescope.

\end{enumerate}

\begin{acknowledgements}
LC acknowledge support from CNPq special visitor fellowship PVE 313945/2013-6 under the Brazilian program Science without Borders. LC, JP, and SA are supported by grants AYA2012-32295 and AYA2012-39408 from the Ministerio de Econom\'ia y Competitividad of Spain. ARA thanks CNPq for partial support  through grant 307403/2012-2. AAH was partially supported by grant AYA2012-31447 from the Ministerio de Econom\'ia y Competitividad of Spain.
\end{acknowledgements}

\begin{landscape}
\begin{table}
\caption{[FeII]1.64$\mu$m/Br$\gamma$ line ratios for LIRGs. Values for emission line peaks and integrated measurements}
\centering
{\small
{\setlength{\tabcolsep}{5pt}
\begin{tabular}{cx{0.7cm}@{ $\pm$ }z{0.7cm}x{0.7cm}@{ $\pm$ }z{0.7cm}x{0.7cm}@{ $\pm$ }z{0.7cm}x{0.7cm}@{ $\pm$ }z{0.7cm}x{0.7cm}@{ $\pm$ }z{0.7cm}x{0.7cm}@{ $\pm$ }z{0.7cm}x{0.7cm}@{ $\pm$ }z{0.7cm}x{0.7cm}@{ $\pm$ }z{0.7cm}x{0.7cm}@{ $\pm$ }z{0.7cm}x{0.7cm}@{ $\pm$ }l}
\hline
\hline
\noalign{\smallskip}
  Object & \multicolumn{4}{c}{Integrated} & \multicolumn{4}{c}{Nucleus} & \multicolumn{4}{c}{HII peak} &\multicolumn{4}{c}{H$_2$ peak} & \multicolumn{4}{c}{[FeII] peak} \\
   & \multicolumn{2}{c}{Obs} & \multicolumn{2}{c}{Corr} & \multicolumn{2}{c}{Obs} & \multicolumn{2}{c}{Corr} & \multicolumn{2}{c}{Obs} & \multicolumn{2}{c}{Corr} & \multicolumn{2}{c}{Obs} & \multicolumn{2}{c}{Corr} & \multicolumn{2}{c}{Obs} & \multicolumn{2}{c}{Corr}  \\
  \noalign{\smallskip}
\noalign{\smallskip}
\hline
NGC2369$^{1,2}$ &   1.26 &   0.22 &   5.33 &   5.27 &   0.81 &   0.06 &   2.49 &   0.98 &   0.83 &   0.06 &   2.42 &   0.89 &   0.84 &   0.06 &   2.49 &   0.93 &   1.16 &   0.10 &   4.72 &   2.35  \\
NGC3110$^{2,3}$ &   0.76 &   0.09 &   3.44 &   1.38 &   0.76 &   0.09 &   5.90 &   5.97 &   0.56 &   0.05 &   0.99 &   0.31 &   0.76 &   0.09 &   6.14 &   6.21 &   0.75 &   0.09 &   4.76 &  13.03  \\
NGC3256 &  \multicolumn{2}{c}{\nodata} &  \multicolumn{2}{c}{\nodata} &  \multicolumn{2}{c}{\nodata} &  \multicolumn{2}{c}{\nodata} &  \multicolumn{2}{c}{\nodata} &  \multicolumn{2}{c}{\nodata} &  \multicolumn{2}{c}{\nodata} &  \multicolumn{2}{c}{\nodata} &  \multicolumn{2}{c}{\nodata} &  \multicolumn{2}{c}{\nodata}  \\
ESO320-G030 &   0.95 &   0.11 &   2.36 &   1.11 &   1.34 &   0.60 &   3.32 &   2.12 &   0.74 &   0.06 &   1.57 &   0.53 &   1.31 &   0.68 &   1.96 &   2.20 &   1.30 &   0.54 &   3.21 &   1.98  \\
IRASF12115-4656$^{1,2}$ & \multicolumn{2}{c}{\nodata} &  \multicolumn{2}{c}{\nodata} &  \multicolumn{2}{c}{\nodata} &  \multicolumn{2}{c}{\nodata} &  \multicolumn{2}{c}{\nodata} &  \multicolumn{2}{c}{\nodata} &  \multicolumn{2}{c}{\nodata} &  \multicolumn{2}{c}{\nodata} &  \multicolumn{2}{c}{\nodata} &  \multicolumn{2}{c}{\nodata}  \\
NGC5135$^{2}$ &   1.10 &   0.10 &   2.79 &   1.27 &   1.03 &   0.06 &   1.86 &   0.67 &   0.74 &   0.03 &   1.43 &   0.34 &   1.03 &   0.06 &   1.86 &   0.67 &   3.74 &   0.15 &   7.71 &   3.09  \\
IRASF17138-1017$^{2}$ &   0.92 &   0.04 &   1.58 &   0.42 &   1.03 &   0.03 &   1.92 &   0.85 &   0.55 &   0.01 &   0.69 &   0.11 &   1.03 &   0.03 &   1.92 &   0.85 &   0.61 &   0.01 &   0.98 &   0.16  \\
IC4687$^{4}$ &   0.85 &   0.04 &   1.23 &   0.18 &   0.93 &   0.06 &   2.07 &   0.63 &   0.43 &   0.02 &   0.61 &   0.07 &   1.05 &   0.08 &   2.08 &   0.72 &   0.47 &   0.02 &   0.66 &   0.08  \\
NGC7130$^{1,2,3}$ &   1.31 &   0.09 &   3.77 &   3.65 &   2.44 &   0.11 &   5.14 &   3.90 &   2.40 &   0.11 &   5.11 &   3.87 &   2.48 &   0.11 &   5.20 &   4.01 &   2.44 &   0.11 &   5.14 &   3.90  \\
IC5179$^{1,2,3}$ &   0.62 &   0.06 &   0.97 &   0.38 &   0.70 &   0.05 &   1.20 &   0.23 &   0.68 &   0.05 &   1.10 &   0.21 &   0.70 &   0.05 &   1.14 &   0.22 &   0.68 &   0.05 &   1.08 &   0.20  \\
\hline
\hline
\end{tabular}}}
\tablefoot{[FeII]/Br$\gamma$ integrated measurements of regions of interest. Observed (obs) and extinction corrected (corr) values are given. $^{1}$ Nucleus and HII peak are coincident. $^{2}$ Nucleus and H$_2$ peak are coincident. $^{3}$ Nucleus and [FeII] peak are coincident. $^{4}$ HII and [FeII] peaks are coincident.}
\label{table:regions_fe_LIRGs}
\end{table}

\begin{table}
\caption{H$_2$ 2.12$\mu$m/Br$\gamma$ line ratios for LIRGs. Values for emission line peaks and integrated measurements}
\centering
{\small
{\setlength{\tabcolsep}{5pt}
\begin{tabular}{cx{0.7cm}@{ $\pm$ }z{0.7cm}x{0.7cm}@{ $\pm$ }z{0.7cm}x{0.7cm}@{ $\pm$ }z{0.7cm}x{0.7cm}@{ $\pm$ }z{0.7cm}x{0.7cm}@{ $\pm$ }z{0.7cm}x{0.7cm}@{ $\pm$ }z{0.7cm}x{0.7cm}@{ $\pm$ }z{0.7cm}x{0.7cm}@{ $\pm$ }z{0.7cm}x{0.7cm}@{ $\pm$ }z{0.7cm}x{0.7cm}@{ $\pm$ }l}
\hline
\hline
\noalign{\smallskip}
  Object & \multicolumn{4}{c}{Integrated} & \multicolumn{4}{c}{Nucleus} & \multicolumn{4}{c}{HII peak} &\multicolumn{4}{c}{H$_2$ peak} & \multicolumn{4}{c}{[FeII] peak} \\
   & \multicolumn{2}{c}{Obs} & \multicolumn{2}{c}{Corr} & \multicolumn{2}{c}{Obs} & \multicolumn{2}{c}{Corr} & \multicolumn{2}{c}{Obs} & \multicolumn{2}{c}{Corr} & \multicolumn{2}{c}{Obs} & \multicolumn{2}{c}{Corr} & \multicolumn{2}{c}{Obs} & \multicolumn{2}{c}{Corr}  \\
  \noalign{\smallskip}
\noalign{\smallskip}
\hline
NGC2369$^{1,2}$ &   1.10 &   0.14 &   1.21 &   0.78 &   0.72 &   0.04 &   0.77 &   0.20 &   0.72 &   0.04 &   0.77 &   0.18 &   0.73 &   0.04 &   0.78 &   0.19 &   0.95 &   0.04 &   1.04 &   0.33  \\
NGC3110$^{2,3}$ &   0.82 &   0.04 &   0.90 &   0.23 &   0.72 &   0.04 &   0.82 &   0.53 &   0.62 &   0.02 &   0.65 &   0.13 &   0.71 &   0.04 &   0.82 &   0.53 &   0.70 &   0.04 &   0.79 &   1.39  \\
NGC3256 &  \multicolumn{2}{c}{\nodata} &  \multicolumn{2}{c}{\nodata} &  \multicolumn{2}{c}{\nodata} &  \multicolumn{2}{c}{\nodata} &  \multicolumn{2}{c}{\nodata} &  \multicolumn{2}{c}{\nodata} &  \multicolumn{2}{c}{\nodata} &  \multicolumn{2}{c}{\nodata} &  \multicolumn{2}{c}{\nodata} &  \multicolumn{2}{c}{\nodata}  \\
ESO320-G030 &   0.88 &   0.05 &   0.93 &   0.28 &   5.47 &   0.46 &   5.81 &   1.77 &   0.46 &   0.03 &   0.49 &   0.11 &   5.90 &   0.55 &   6.26 &   1.92 &   4.74 &   0.36 &   5.03 &   1.52  \\
IRASF12115-4656$^{1,2}$ &   0.84 &   0.08 &   0.95 &   0.84 &   1.69 &   0.19 &   1.92 &   1.70 &   1.66 &   0.19 &   1.88 &   1.67 &   1.69 &   0.19 &   1.92 &   1.70 &   \multicolumn{2}{c}{\nodata} &   \multicolumn{2}{c}{\nodata}  \\
NGC5135$^{2}$ &   1.29 &   0.07 &   1.37 &   0.40 &   1.70 &   0.06 &   1.76 &   0.41 &   0.60 &   0.02 &   0.62 &   0.10 &   1.70 &   0.06 &   1.76 &   0.41 &   1.62 &   0.08 &   1.70 &   0.44  \\
IRASF17138-1017$^{2}$ &   0.47 &   0.02 &   0.49 &   0.08 &   0.49 &   0.02 &   0.52 &   0.15 &   0.16 &   0.01 &   0.17 &   0.02 &   0.49 &   0.02 &   0.52 &   0.15 &   0.26 &   0.01 &   0.27 &   0.03  \\
IC4687$^{4}$ &   0.38 &   0.01 &   0.39 &   0.04 &   0.56 &   0.02 &   0.59 &   0.12 &   0.13 &   0.01 &   0.14 &   0.01 &   0.68 &   0.03 &   0.71 &   0.16 &   0.14 &   0.01 &   0.15 &   0.01  \\
NGC7130$^{1,2,3}$ &   1.41 &   0.07 &   1.51 &   0.94 &   1.35 &   0.04 &   1.42 &   0.69 &   1.34 &   0.04 &   1.40 &   0.68 &   1.37 &   0.04 &   1.44 &   0.72 &   1.35 &   0.04 &   1.42 &   0.69  \\
IC5179$^{1,2,3}$ &   0.66 &   0.03 &   0.68 &   0.17 &   0.49 &   0.02 &   0.51 &   0.06 &   0.47 &   0.02 &   0.49 &   0.06 &   0.49 &   0.02 &   0.50 &   0.06 &   0.47 &   0.02 &   0.48 &   0.06  \\
\hline
\hline
\end{tabular}}}
\tablefoot{H$_2$ 2.12$\mu$m/Br$\gamma$ integrated measurements of regions of interest. Observed (obs) and extinction corrected (corr) values are given. $^{1}$ Nucleus and HII peak are coincident. $^{2}$ Nucleus and H$_2$ peak are coincident. $^{3}$ Nucleus and [FeII] peak are coincident. $^{4}$ HII and [FeII] peaks are coincident.}
\label{table:regions_h2_LIRGs}
\end{table}

\end{landscape}

-------------------------------------------------------------------

\section{Appendix}

\begin{figure*}
\centering
\includegraphics[width=16cm]{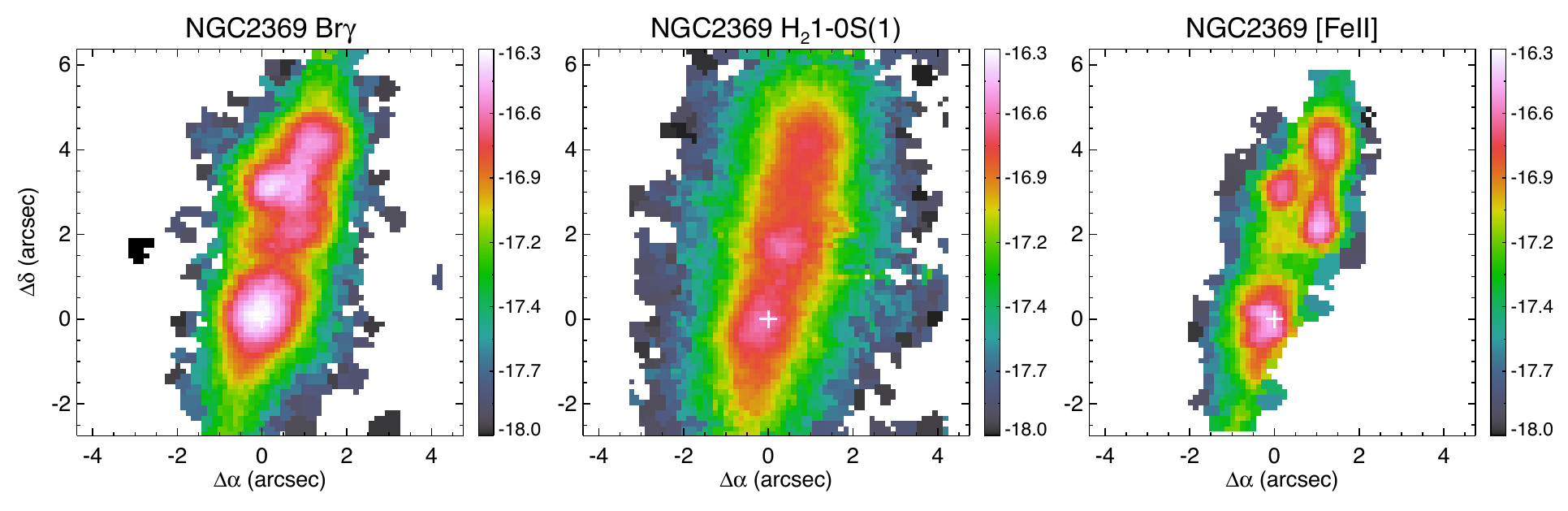}
\includegraphics[width=16cm]{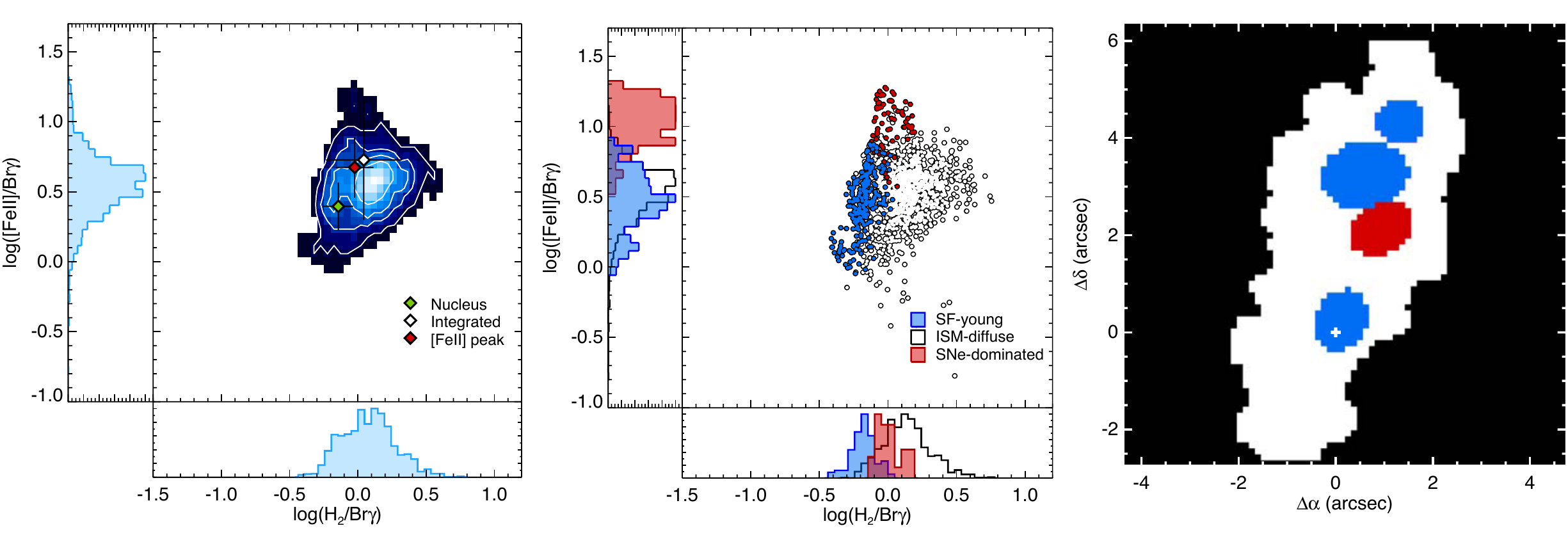}
\caption{Emission line maps and near-IR diagnostic planes for NGC2369. Colour-coded symbols as in Fig.~\ref{figure:IC4687}}
\label{figure:NGC2369}
\end{figure*}

\begin{figure*}
\centering
\includegraphics[width=16.5cm]{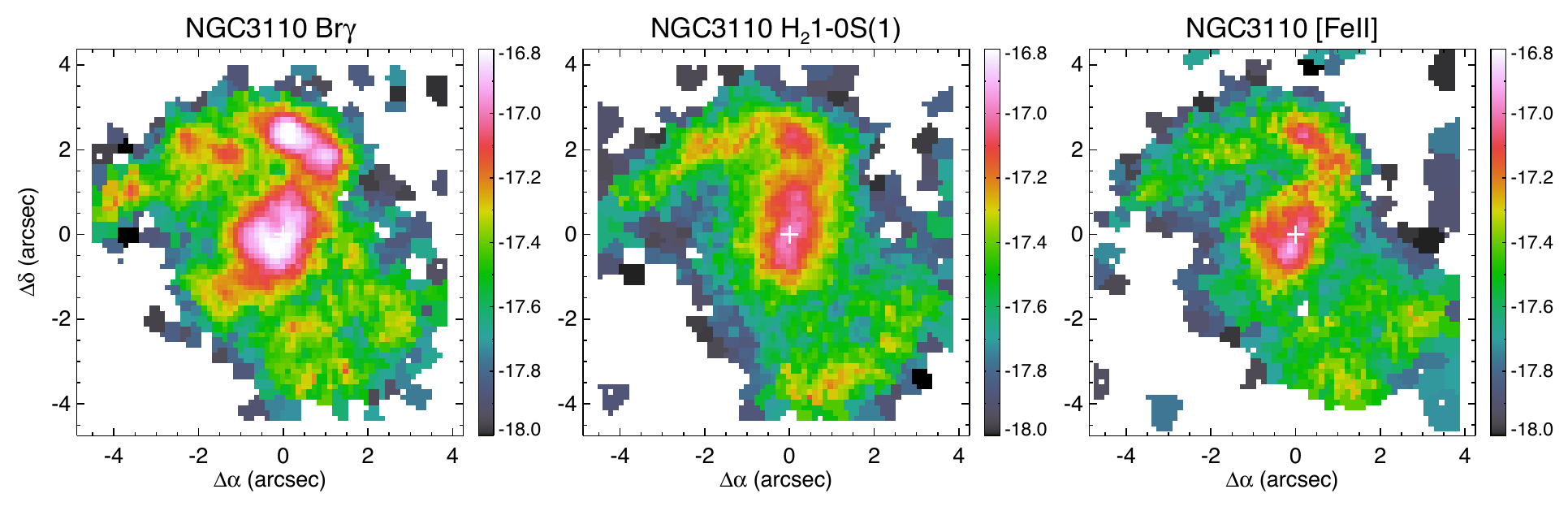}
\includegraphics[width=16.5cm]{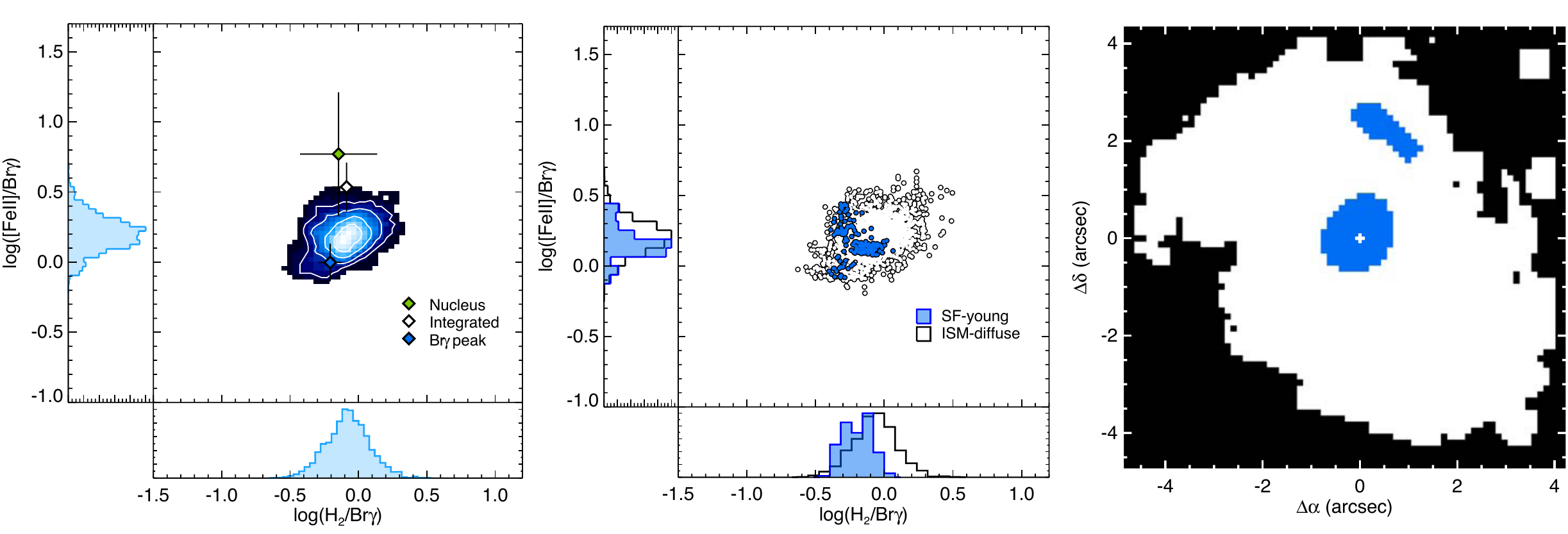}
\caption{Emission line maps and near-IR diagnostic planes for NGC3110. Colour-coded symbols as in Fig.~\ref{figure:IC4687}}
\label{figure:NGC3110}
\end{figure*}

\begin{figure*}
\centering
\includegraphics[width=18.5cm]{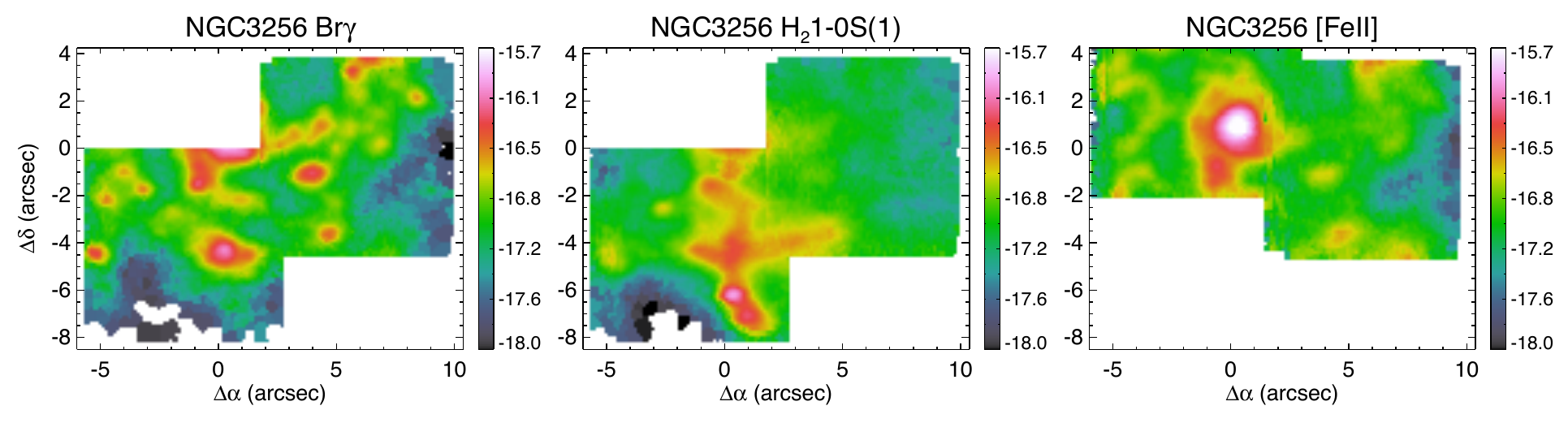}
\includegraphics[width=18.5cm]{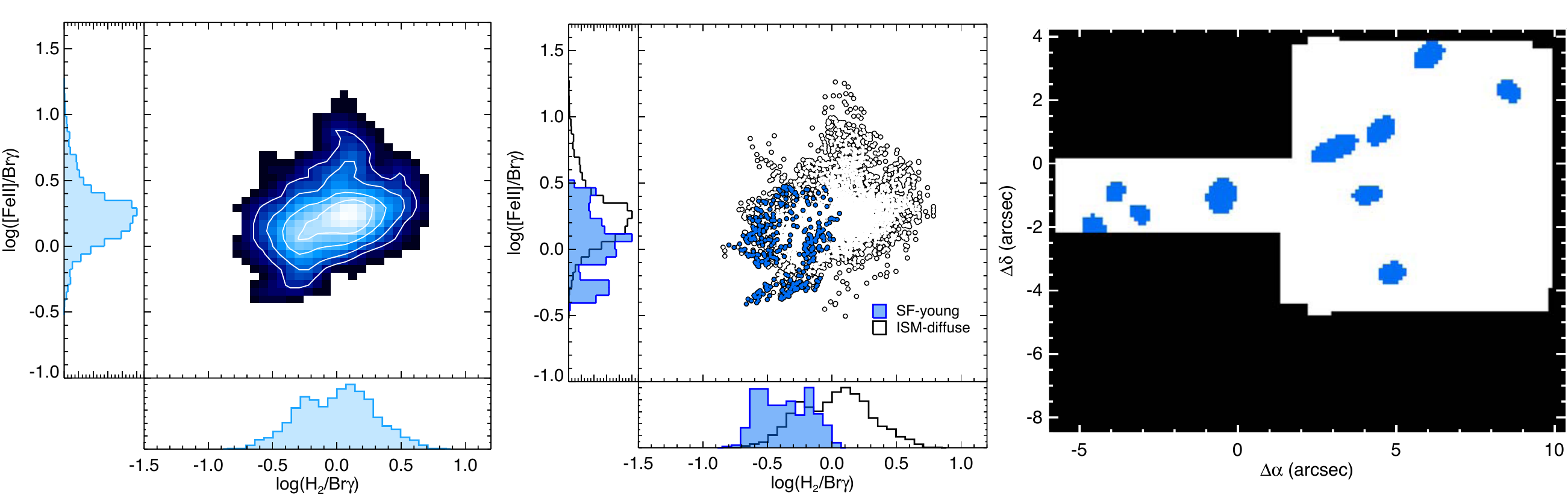}
\caption{Emission line maps and near-IR diagnostic planes for NGC3256. Colour-coded symbols as in Fig.~\ref{figure:IC4687}}
\label{figure:NGC3256}
\end{figure*}

\begin{figure*}
\centering
\includegraphics[width=17cm]{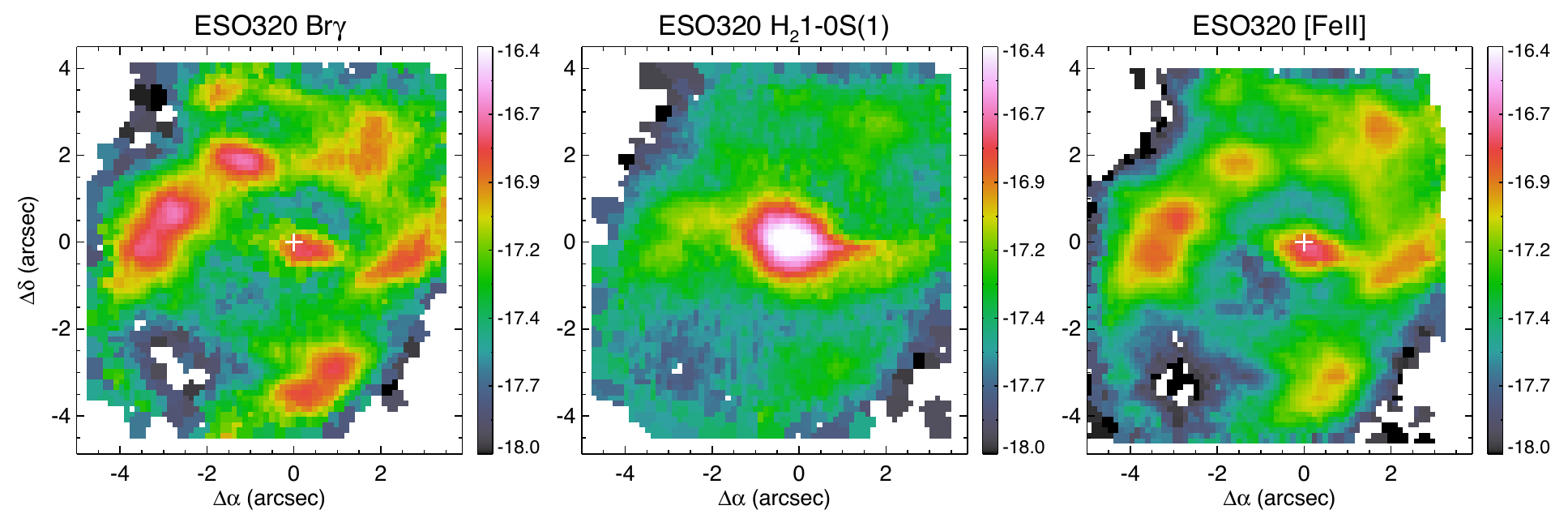}
\includegraphics[width=17cm]{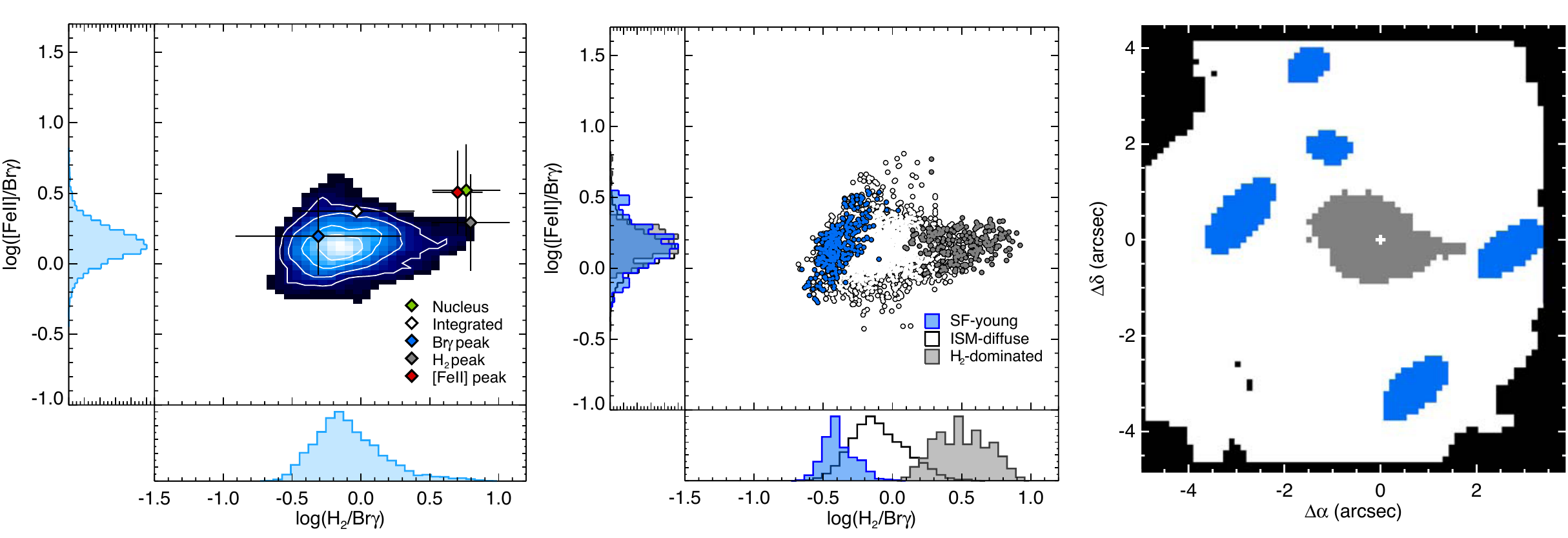}
\caption{Emission line maps and near-IR diagnostic planes for ESO320$-$G030. Colour-coded symbols as in Fig.~\ref{figure:IC4687}}
\label{figure:ESO320}
\end{figure*}

\begin{figure*}
\centering
\includegraphics[width=16cm]{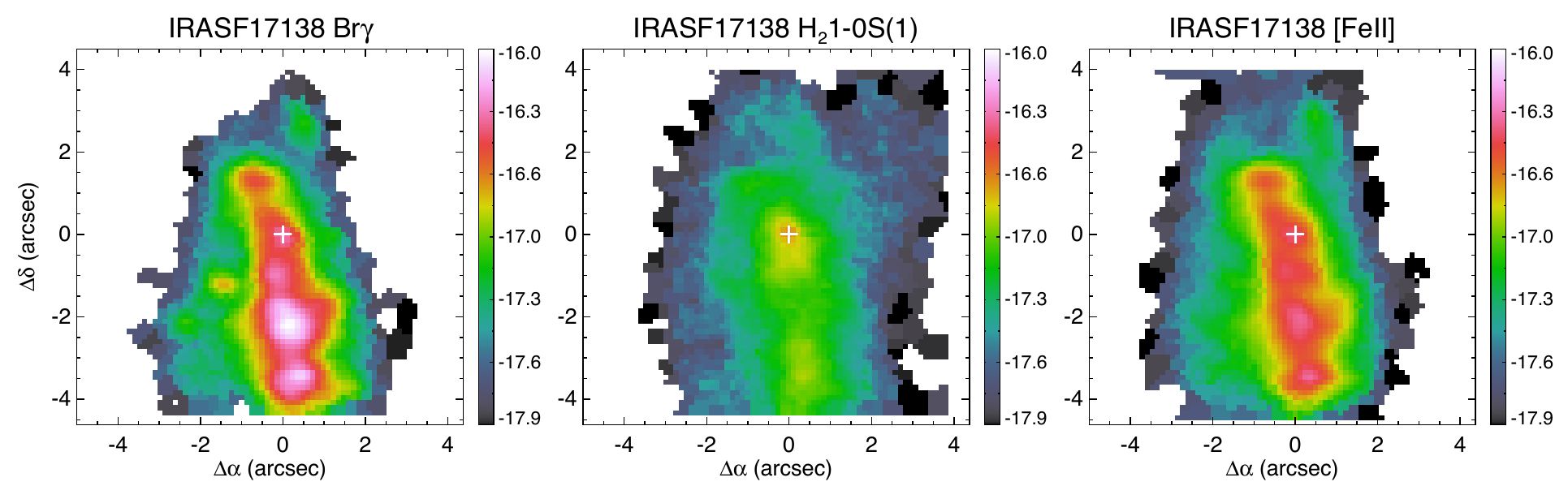}
\includegraphics[width=16cm]{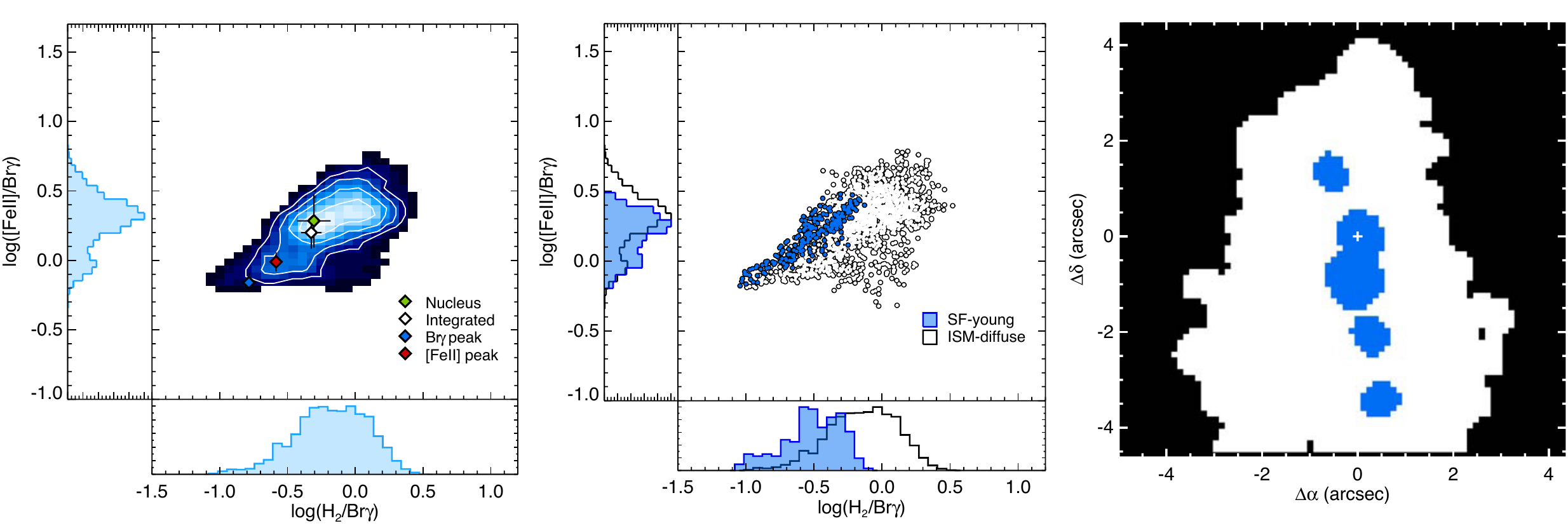}
\caption{Emission line maps and near-IR diagnostic planes for IRASF17138$-$1017. Colour-coded symbols as in Fig.~\ref{figure:IC4687}}
\label{figure:IRASF17138}
\end{figure*}

\begin{figure*}
\centering
\includegraphics[width=16cm]{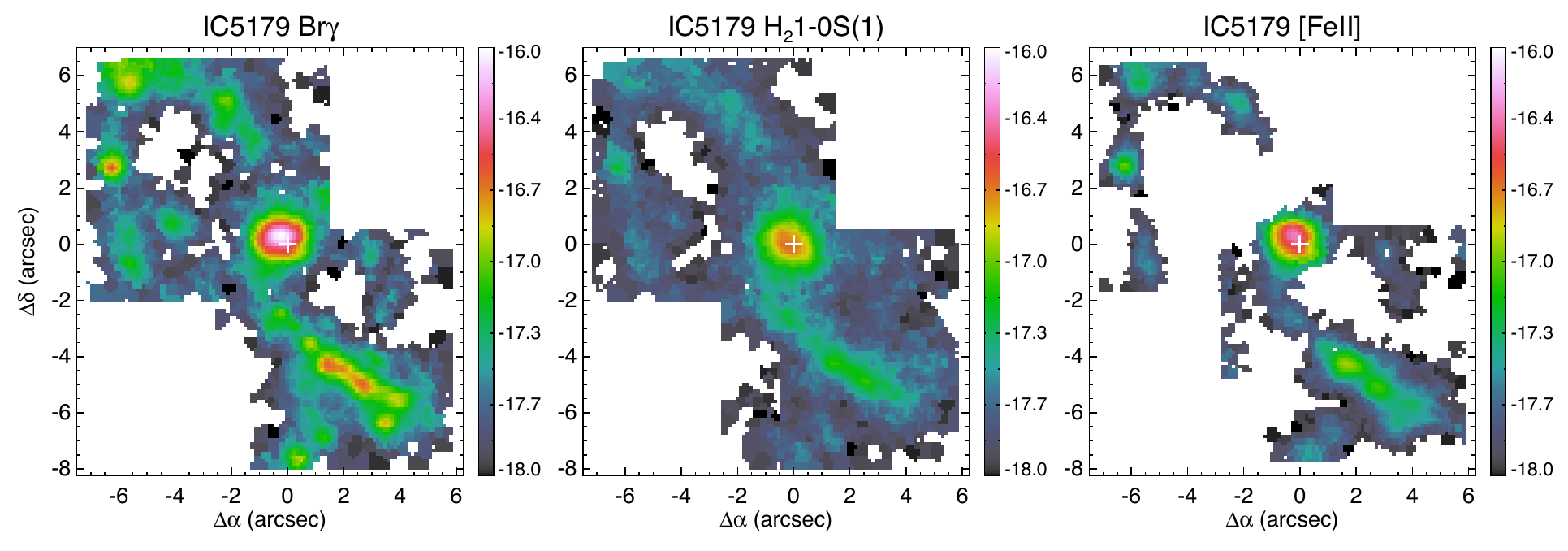}
\includegraphics[width=16cm]{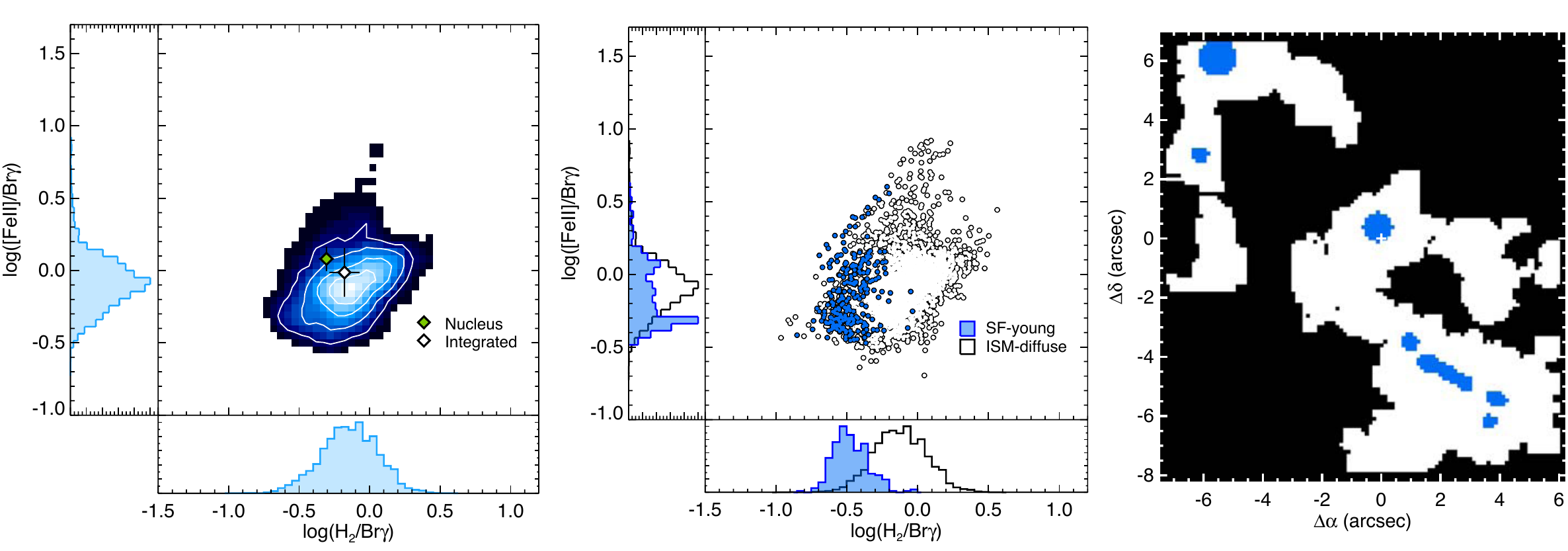}
\caption{Emission line maps and near-IR diagnostic planes for IC5179. Colour-coded symbols as in Fig.~\ref{figure:IC4687}}
\label{figure:IC5179}
\end{figure*}

The appendix presents in graphical form (Figs.~\ref{figure:NGC2369} to \ref{figure:IC5179}), the emission line maps and distribution of the emission line ratios in the log(FeII]1.64$\mu$m/Br$\gamma$) $-$ log(H$_2$2.12$\mu$m/Br$\gamma$) plane for the sample of LIRGs, excluding the prototypes that have already been discussed (Figs.~\ref{figure:IC4687} to \ref{figure:NGC5135}). Notes on individual galaxies can be found elsewhere \citep{Piqueras2012A&A546A}

\bibliographystyle{aa}
\bibliography{aa-SINFONI-BPT}

\end{document}